\documentclass[prb,twocolumn,preprintnumbers,amsmath,amssymb]{revtex4-1}
\usepackage{graphicx}

\begin{document}
\title{ Fermionic Hubbard model with Rashba or Dresselhaus spin-orbit coupling }
\author{ Fadi Sun$^{1,2}$, Jinwu Ye$^{1,3,4}$, and Wu-Ming Liu$^{2}$  }
\affiliation{
$^{1}$Department of Physics and Astronomy, Mississippi State University, MS, 39762, USA \\
$^{2}$Beijing National Laboratory for Condensed Matter Physics, Institute of Physics, Chinese Academy of Sciences, Beijing 100190, China   \\
$^{3}$ Department of Physics, Capital Normal University,
Key Laboratory of Terahertz Optoelectronics, Ministry of Education, and Beijing Advanced innovation Center for Imaging Technology,
Beijing, 100048, China   \\
$^{4}$  Kavli Institute of Theoretical Physics, University of California, Santa Barbara, Santa Barbara, CA 93106  }

\date{\today }

\begin{abstract}
In this work, we investigate the possible dramatic effects of Rashba or Dresselhaus spin-orbit coupling (SOC) on fermionic Hubbard model
in a 2d square lattice.
In the strong coupling limit, it leads to the Rotated Anti-ferromagnetic Heisenberg model
which is a new class of quantum spin model.
For a special equivalent class,
we identify a new spin-orbital entangled commensurate ground ( Y-y ) state subject to strong quantum fluctuations at $T=0$.
We evaluate the quantum fluctuations by the spin wave expansion up to order $ 1/S^2 $.
In some SOC parameter regime, the Y-y state supports a massive relativistic in-commensurate magnon ( C-IC )
with its two gap minima positions continuously tuned by the SOC parameters.
The C-IC magnons dominate all the low temperature thermodynamic quantities and also
lead to the separation of the peak positions between the longitudinal and the transverse spin structure factors.
In the weak coupling limit, any weak repulsive interaction also leads to a weak Y-y state.
There is only a crossover from the weak to the strong coupling.
High temperature expansions of the specific heats in both weak and strong coupling are presented.
The dramatic roles to be played by these C-IC magnons at generic SOC parameters or under various external probes are hinted.
Experimental applications to both layered noncentrosymmetric materials and cold atom are discussed.
\end{abstract}

\maketitle


\section{Introduction }

 It was well known that it is the strong electron correlations \cite{wen0,scaling,aue,wenbook,sachdev,stevermp}
 which lead to many important phenomena such as
 Anti-ferromagnetism, spin density wave, charge density wave, putative spin liquids with topological orders, un-conventional superconductivity,etc.
 The Rashba or Dresselhaus spin-orbit coupling (SOC)  \cite{rashba} is ubiquitous in various
 2d or layered insulators, semi-conductor systems,
 metals and superconductors without inversion symmetry \cite{ahe,socsemi,niu,ahe2,she,aherev,sherev}.
 On the other forefront, due to their tunability and controllability, strongly correlated Fermi gases on optical lattices
 have been attempted with some success to quantum simulate some of these phenomena \cite{blochrmp,coldafm}.
 There are  very recent notable experimental advances in generating 2d Rashba or Dresselhaus SOC or any their linear combinations
 for cold atoms in both continuum and optical lattices \cite{expk40,expk40zeeman,clock,2dsocbec}.
 It becomes topical and important to investigate the combined effects of strong correlations and Rashba SOC in various lattice systems.


 In this paper, we address this outstanding problem.
 Specifically,  we investigate the system of interacting fermions
 at half filling hopping in a 2 dimensional square lattice subject to any combinations of Rashba or Dresselhaus SOC.
 In the strong coupling limit, we reach a novel quantum spin model named Rotated Anti-Ferromagnetic Heisenberg
 model (RAFHM) which is a new class of quantum spin models.
 For a special combination of the Rashba or Dresselhaus SOC, we identify a new spin-orbital entangled commensurate ground
 called Y-y state \cite{notation} subject to quantum fluctuations at $T=0$.
 We evaluate the quantum fluctuations by spin wave expansions up to $ 1/S^2 $ order
 ( which is also called $ 1/S $ correction to the linear spin wave expansion  ( LSWE ) in previous literatures on Heisenberg models ).
It supports a massive relativistic  commensurate magnon $ C-C_0 $ in one SOC parameter regime
and an in-commensurate magnon C-IC  in the other regime \cite{notation}.
The two gap minima positions of the C-IC magnons are continuously tuned by the SOC strength.
At low temperatures, these magnons dominate all the physical quantities such as the specific heat, magnetization,
$ (0,\pi) $ and $ (\pi,0) $ susceptibilities, Wilson ratio and also various spin correlation functions.
At $ T=0 $, the longitudinal spin structure factor  shows a sharp peak at $ \vec{k}=0 $ in the reduced Brillouin Zone (RBZ)
reflecting the ground state.
However, the transverse spin structure factor displays non-trivial features reflecting the magnon excitations above the ground state.
The $ C-C_0 $ leads to a pinned central Lorentzian peak at $ \vec{k}=0 $
in the TSSF. However, the $ C-IC $ splits it into two Lorentzian peaks
located at its two gap minima, while changing its structure at $\mathbf{k}=0 $ into a saddle point one.
In the weak coupling limit, any weak repulsive interaction leads to a weak Y-y state which also hosts low energy fermionic excitations.
There is a crossover from the weak to the strong coupling where the fermionic excitation energies increase.
The electronic and spin Wilson loops can be determined by measuring specific heats in high temperature expansion in weak
and strong coupling limit respectively.
The $ C-IC $ encodes short-range incommensurate seeds embedded in an commensurate ground state at $ T=0 $, which justifies its name \cite{notation}.
The crucial roles to be played by these seeds at generic SOC parameters $ (\alpha,\beta) $ and
under various external probes are outlined in the conclusion section.
Experimental realizations and detections in both layered noncentrosymmetric materials and cold atom systems are discussed.

\section{ The interacting fermionic model and the quantum spin model in the strong coupling limit:}
\begin{figure}
\includegraphics[width=0.98\linewidth]{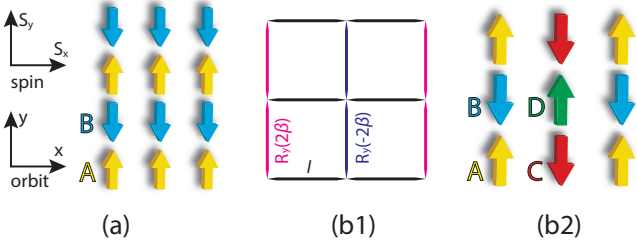}
\caption{ (Color Online) (a) The Y-y  ground state in a square lattice in the original basis.
 One only need to introduce 2 HP bosons corresponding to the $ A, B $ sublattice structure.
(b1) The $ (\pi,0) $ sublattice structure of the Hamiltonian Eqn.\ref{rh} in the $ U(1) $ basis.
(b2) The $ Y-(\pi,\pi) $ Neel state in the $ U(1) $ basis.
 Due to the in-compatibility of the two sublattice structures in (b), one need to introduce four
 HP bosons corresponding to the 4-sublattice structures $ A,B,C,D $ to perform the SWE.
 Due to the 4-sublattice structure, the Reduced Brillouin Zone ( RBZ ) is 4 times smaller than the full BZ.}
\label{u1}
\end{figure}

The tight-binding Hamiltonian of spin $ 1/2 $ fermions at half filling  hopping in
a two-dimensional (2D) square optical lattice subject to any combination of Rashba and Dresselhaus SOC is:
\begin{equation}
	\mathcal{H}_f= -t\sum_{\langle ij\rangle}(c_{i\sigma}^\dagger U_{ij}^{\sigma\sigma'} c_{j\sigma'}+h.c.)	+ U\sum_i(n_{i\uparrow}-\frac{1}{2})(n_{i\downarrow}-\frac{1}{2})
\label{hubbardint}
\end{equation}
where $ t $ is the hopping amplitude along the nearest neighbors $\langle ij\rangle$,
the non-Abelian gauge fields $U_{iㄛi+\hat{x} } =e^{i \alpha \sigma_x}$, $U_{iㄛi+\hat{y} }=e^{i \beta \sigma_y}$
are put on the two links in Fig.\ref{u1}a which is  the lattice regularization of
the linear combination $ k_x \sigma_x + k_y \sigma_y $ of the Rashba  and Dresselhaus SOC in a continuum momentum space \cite{ahe,she}.
$ \alpha=\pm \beta $ stands for the isotropic Rashba ( Dresselhaus ) case.
$U>0$ is the Hubbard onsite interaction.

In the strong coupling limit $ U/t \gg 1 $, to the order $O(t^2/U)$,
we obtain the effective spin $ 1/2 $ Rotated anti-ferromagnetic Heisenberg model (RAFHM):
\begin{align}
	\mathcal{H}_{RH}=J\sum_i
	[S^a_i R_{ab}(X,2\alpha) S^b_{i+\hat{x}}
	+S^a_i R_{ab}(Y,2\beta) S^b_{i+\hat{y}}]
\label{rh}
\end{align}
with  the anti-ferromagnetic exchange interaction $J=4t^2/U > 0 $,
$a, b=1,2,3 $ are the three components of the spin operator,
the $R(X,2\alpha)$, $R(Y,2\beta)$ are the two SO(3) rotation matrices
around the $ X $ and $ Y $ spin axis by angle $2\alpha$, $2\beta$
putting on the two bonds along  $\hat{x} $, $\hat{y} $ respectively.

Here, we plan to study the quantum  or topological phenomena in the RH model at generic $(\alpha,\beta)$.
However, it is a very difficult task, so we take a "divide and conquer" strategy.
First, we identify a solvable line  $(\alpha=\pi/2, \beta)$ and explore new and rich quantum phenomena along the line.
Then starting from the results achieved from the solvable line, then we will investigate the quantum phenomena
at the generic $ ( \alpha, \beta) $ including the Rashba or  Dresselhaus SOC point $ \alpha= \pm \beta $.
In this paper, we will focus on the first task.
The second task will be outlined in the conclusion section and presented in details elsewhere.
In the past, this kind of ``divide and conquer'' approach has been very successful
in solving many quantum spin models.
For example, in the single ( multi-) channel Kondo model, one solve the Thouless  (Emery-Kivelson) line \cite{kondo1,kondo2}, then do perturbation away from it.
In quantum-dimer model, one solves the Rohksa-Kivelson (RK) point which shows spin liquid physics \cite{dimer}, then one can study the effects of various
perturbations away from it \cite{dimer2}. Recently, this "divide and conquer" strategy was quite successfully applied to
study the RFHM along the solvable line first in \cite{rh}, then at the generic SOC parameter in \cite{rhrashba}.

\begin{figure}
\includegraphics[width=0.98\linewidth]{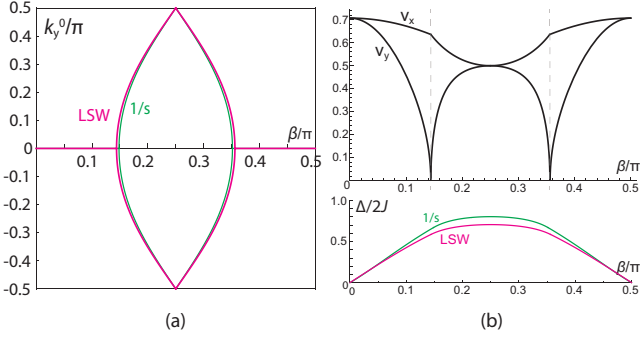}
\caption{ (Color Online)
(a) The minima position $\mathbf{k}=(0, \pm k_y^0)$ of the relativistic magnons in the reduced Brillouin Zone (RBZ).
(b) The energy gap  ( or mass ) $\Delta(\beta)$ at the minima in (a) with the two magnon velocities
    $ v_x \geq v_y $. The equality holds at $ \beta=0, \pi/4, \pi/2 $.
    Near $ \beta_i, i=1,2 $, $ v_x $ has a cusp, while $ v_y \sim |\beta-\beta_i|^{1/2} $.
The LSWE ( $ 1/S $ order ) results are in purple line and $ 1/S $ corrections to LSWE in green line.
The $1/S $ corrections are found to be small ( see the appendix B).  }
\label{groundex}
\end{figure}

The RAFHM Eqn.\ref{rh} inherits all the symmetries of the fermionic model Eqn.\ref{hubbardint}.
Along the line $(\alpha=\pi/2, \beta)$, in addition to own the spin-orbital coupled $ U(1)_{soc} $ symmetry $[H_f,\sum_i(-1)^{i_x}c_i^\dagger\sigma^y c_i]=0 $,
it also has an extra mirror $ {\cal M} $ symmetry:
under the local rotation $ \tilde{\mathbf{S}}_{i} =R(\hat{x},\pi ) R(\hat{y},\pi n_2) \mathbf{S}_{i}$, then followed by a Time reversal transformation,
$ \beta \rightarrow \pi/2 - \beta $. At the middle point $ \beta=\pi/4 $, the Hamiltonian is invariant under such a Mirror transformation.
In the classical limit $S\to\infty$, one can show that the ground state is the $Y-y$ state in Fig.\ref{u1}a which still respects
both the $ U(1)_{soc} $ symmetry and the $ {\cal M} $ symmetry
which will be used to classify the symmetry of the minimum positions
and the magnon gap in Fig.\ref{groundex}. The $ Y-y $ state also keeps the
$ {\cal P}_y $ and $  {\cal T} {\cal P}_x $ and $  {\cal T} {\cal P}_z $ symmetries.

\begin{figure}
\includegraphics[width=0.98\linewidth]{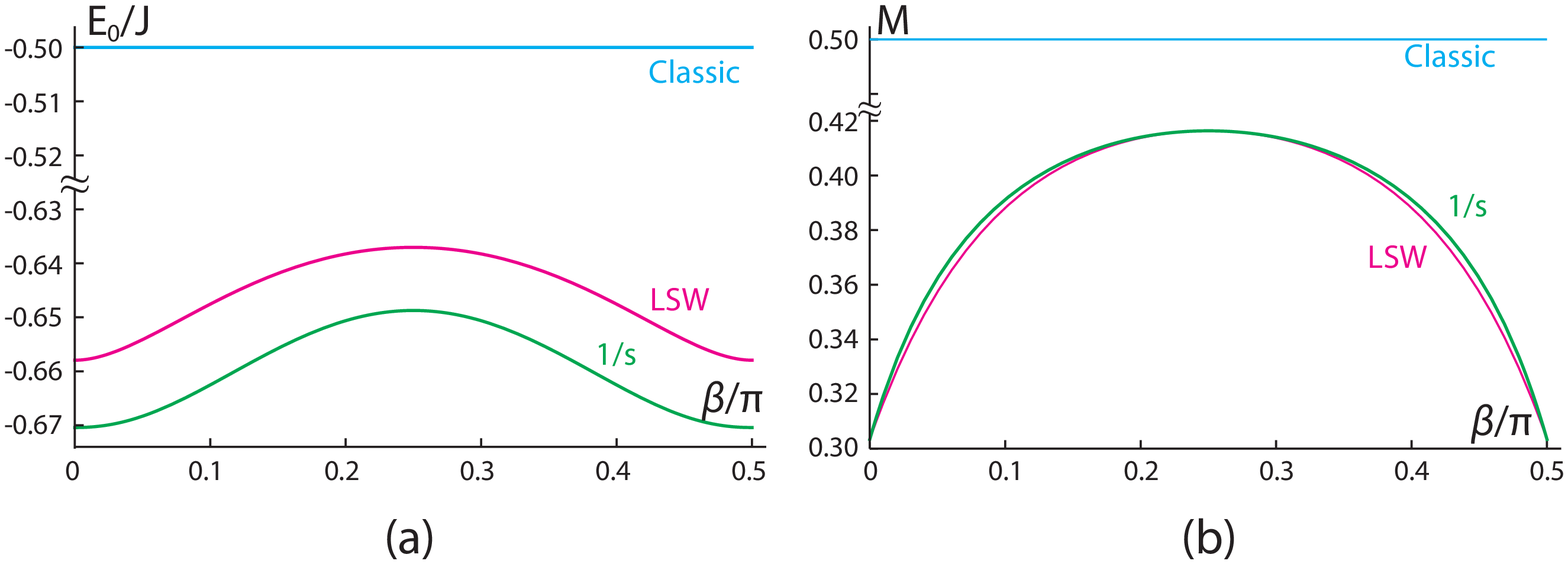}
\caption{ (a) The ground-state energy and (b) the magnetization
as a function of gauge field parameter $ 0 < \beta < \pi/2 $ along the line. Shown are the classical results in blue color( flat line on top )
which  are independent of $ \beta $, LSWE ( $ 1/S $ order ) in purple line and $ 1/S $ corrections to LSWE in green line.
There are always $ 1/S $ corrections to the ground state energy in (a).
In (b), it vanishes at the two Abelian points $ \beta=0,\pi $ and also at $ \beta=\pi/4 $.
The $1/S $ corrections are found to be small in both quantities. This fact shows that the LSW is quite accurate even for the smallest $ s=1/2 $
which hosts the largest quantum fluctuations. }
\label{eM1s}
\end{figure}

\section{ The $ C-C_0 $ and $ C-IC $ magnons above the $ Y-y $ state.}
Based on the $Y-y$ state in Fig.\ref{u1}a,
we introduce the Holstein-Primakoff (HP) bosons $ a $ and $ b $ for the sublattice $A$ and
$ B $ respectively, the Hamiltonian Eq.\eqref{rh} can be written in a systematic $ 1/S $ expansion
in terms of the HP bosons \cite{sw1,sw2,japan,higgs}:
\begin{equation}
	\mathcal{H}_{\rm spin}
	=\mathcal{H}_0+2JS(\mathcal{H}_2+\mathcal{H}_4+\cdots)
\end{equation}
where $\mathcal{H}_0=-2NJS^2$ is the classical ground state energy.
The $\mathcal{H}_2$ represents linear spin wave theory,
the $\mathcal{H}_4$ represents $1/S$ correction to linear spin wave theory \cite{japan} and so on.
In the rest of the paper, we will use $ 2 J S $ to be the energy unit.

By combining a unitary transformation, followed by a Bogoliubov transformation ( see the appendix A ),
one can diagonize $\mathcal{H}_2$:
\begin{align}
	\mathcal{H}_2\!=\!\sum_k(\omega_k^+\!+\omega_k^-\!-2)
	  +2\sum_k(\omega_k^-\alpha_k^\dagger\alpha_k
		  +\omega_k^+\beta_k^\dagger\beta_k)
\label{LSW}
\end{align}
where the LSW spectrum $\omega_k^\pm=\sqrt{1-(\gamma_k^\pm)^2}$
and $2\gamma_k^{\pm}=\cos2\beta\cos k_y
\pm\sqrt{\cos^2 k_x+\sin^22\beta\sin^2k_y}$.
When $ \beta < \pi/4 $, $ \omega^{+}_{k} < \omega^{-}_{k} $,
when $ \beta > \pi/4 $, $ \omega^{+}_{k} > \omega^{-}_{k} $,
at $ \beta = \pi/4 $, $ \omega^{+}_{k} = \omega^{-}_{k} $.
So $ \omega^{+}_{k} $ and $ \omega^{-}_{k} $ are related by the $ {\cal M} $ symmetry.
In the following, for the notational simplicity, we call the lower branch  $ \omega^{-}_{k} $,  the energy is measured in the unit of $4JS$.

Along the line $ (\alpha=\pi/2, 0 < \beta < \pi/2 ) $,
the position of the minima of the lower branch $\omega_k^{-} $ is given in Eq.\ref{k0y} and shown in Fig.\ref{groundex}a.
One can see that  when $ 0 < \beta < \beta_1 $ and $ \beta_2=\pi/2-\beta_1 < \beta < \pi/2 $, the $ Y-y $ ground state supports the
C-C$_0$, when $ \beta_1 < \beta < \beta_2 $, it supports the C-IC magnons.
The low energy excitation can be obtained from the expansion around the minima
$\mathbf{k}=\mathbf{k}_0+\mathbf{q}$ as:
\begin{equation}
\omega^{-}_q=\sqrt{\Delta^2(\beta)+v_x^2q_x^2+v_y^2q_y^2}
\label{disper}
\end{equation}
 where the mass $ \Delta $ at the minima and the two velocities are given in Eqn.\ref{disper2} and shown in Fig.\ref{groundex}b.

Thus, they are relativistic gapped particles with a gap $\Delta$
and two velocities $v_x \geq v_y$  where the equality holds at $ \beta=0, \pi/4, \pi/2 $.
Near $ \beta_i, i=1,2 $, $ v_y \sim |\beta-\beta_i|^{1/2} $.
In a sharp contrast, the C-IC magnons in the RFHM \cite{rh} are non-relativistic gapped particles
with a gap $\Delta$ and  two effective masses $m_y \geq m_x $.

At the two Abelian points $\beta=0,\pi/2$,
the system has SU(2) symmetry in the rotated basis $ \tilde{SU}(2) $ with $\tilde{\mathbf{S}}_{i} = R(\hat{x},\pi n_1) \mathbf{S}_{i}$
and $ \tilde{\tilde{SU}}(2) $ with $ \tilde{\tilde{\mathbf{S}}}_{i} = R(\hat{x},\pi n_1)  R(\hat{y},\pi n_2) \mathbf{S}_{i} $
respectively ( Fig.\ref{weak} ),
Eq.\ref{LSW} reduces to the AFM spin wave $\omega_k\sim k$ at the minimum $(0,0)$ and $(\pi,0)$ respectively.

 We also obtain the ground-state energy and the magnetization
   at $ T=0 $ from the LSW:
\begin{eqnarray}
	E_{\rm GS}&=&2NJS^2+2JS\sum_k(\omega_k^++\omega_k^--2)   \nonumber  \\
	M&=&S-\frac{1}{2N}\sum_{k} ( \frac{1}{\omega^{+}_{k}} + \frac{1}{\omega^{-}_{k}}-2 )
\label{eM}
\end{eqnarray}
   which are drawn in Fig.\ref{eM1s}.

\section{ Thermodynamic quantities at low temperatures. }

\begin{figure}
\includegraphics[width=0.88\linewidth]{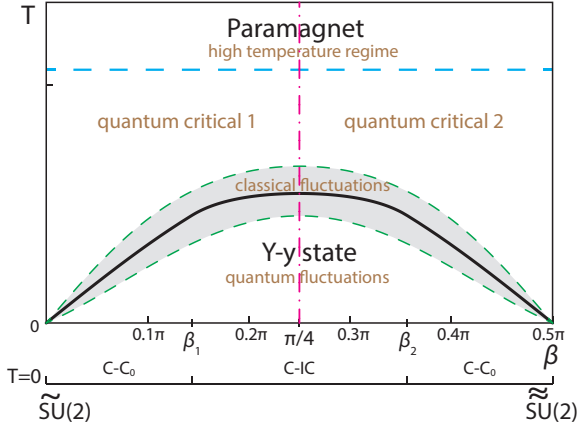}
\caption{ (Color Online)  The classical fluctuations dominate in the narrow regime around the finite temperature transition denoted by the two dashed lines,
so it is still in the 2d Ising transition class as that in the RFHM. The quantum fluctuations dominate in the Y-y state below the dashed line where the Sec.IV focus. The QC1 and QC2 regime where $ \Delta \ll T $
are controlled by the two abelian points at $ \beta=0 $ and $ \beta=\pi/2 $ respectively, so are dominated by the
quantum Anti-ferromagnetic fluctuations  in the $\tilde{\mathbf{SU}}(2) $ and $ \tilde{\tilde{\mathbf{SU}}}(2) $ basis respectively.
The high temperature expansion in the Sec.VII holds only in the high temperature regime. }
\label{finiteT}
\end{figure}

   At low temperatures, one can drop the higher energy mode of the $ \omega^{+}_{k} $ and evaluate the specific heat
   and the Staggered magnetization of the $Y-y$ state in Fig.\ref{finiteT}  due to the relativistic magnons:
\begin{eqnarray}
	C_m(T) &\sim & \frac{\Delta^3}{2\pi v_xv_y T}e^{-\Delta/T}  \nonumber  \\
    M(T) &\sim  & M-\frac{T^2}{2\pi v_xv_y}e^{-\Delta/T}
\label{CmT}
\end{eqnarray}
    where $M $ is the zero temperature staggered magnetization listed in Eq.\ref{eM}.


   By coupling to the conserved quantity $- H_s \sum_i (-1)^{i_x}S_i^y$ and to the order parameter $-H_s\sum_i (-1)^{i_y}S_i^y$
   respectively, one can also evaluate the $(\pi,0)$ and $ (0,\pi) $ staggered susceptibilities:
\begin{eqnarray}
    \chi_{(\pi,0)}(T) & \sim & \frac{\Delta}{2\pi v_xv_y} e^{-\Delta/T}    \nonumber   \\
     \chi_{(0,\pi)}(T) & \sim & \chi_{(0,\pi)}(T=0)-\frac{1}{2\pi v_xv_y}
    \frac{1}{\Delta}e^{-\Delta/T}
\label{pizero}
\end{eqnarray}
   where  $  \chi_{(0,\pi)}(T=0) =\sum_{k,s=\pm} \frac{1-(\omega_k^s)^2}{2(\omega_k^s)^3}$ is the
   zero temperature $(0,\pi)$ staggered susceptibility.



From the specific heat $C_m(T)$ in Eq.\ref{CmT}
and the conserved $ (\pi,0) $ staggered susceptibility $\chi_{(\pi,0)}(T)$ in Eq.\ref{pizero},
one can form the Wilson ratio:
\begin{align}
    R_w=\frac{T\chi_{(\pi,0)}(T)}{C_m(T)}=\Big(\frac{T}{\Delta}\Big)^2
\label{wilson}
\end{align}
which only depends on \cite{setting} the dimensionless quantity of $T/\Delta(\beta)$.

 The physical quantities in Eq.\ref{CmT} and \ref{pizero} depend on explicitly the magnon's two velocities $ v_x, v_y $ and its gap
 $ \Delta $ shown in Fig.\ref{groundex}b. However, the Wilson ratio in Eq.\ref{wilson} only depends on the gap $ \Delta $.
 It is easy to see that the longitudinal spin structure factor always has a very sharp peak $ M^2 \delta_{\vec{k},0} $
 at the ordering wavevector $ (0, \pi) $ ( which is at $ (0,0) $ in the RBZ ) of the Y-y state in Fig.\ref{u1}a with the spectral weight
 equal to the square of the magnetization.
 Unfortunately, the positions of the gap minima at $ (0, k^{0}_y ) $ of the C-IC magnons in Fig.\ref{groundex}a can not be reflected in
 all these physical quantities. In the following, we show that they can be precisely mapped out by the peak positions in
 the transverse structure factors.




\section{ Transverse Structure factors in the $ U(1) $ basis. }
 Performing a local gauge transformation $ \tilde{c}_i=(i\sigma_x)^{i_x}c_i $ on Eqn.\ref{hubbardint}, one can
 get rid of the gauge fields on all the $x-$ links, all the remaining gauge fields on the $ y-$ links  commute.
 Similarly, by performing a local rotation $ \tilde{\mathbf{S}}_n = R( \hat{x}, \pi n_1 ) \mathbf{S}_n $ in Eqn.\ref{rh}, one can get rid of the
 $R$-matrix on the $x$-links shown in Fig.\ref{u1}b1. It makes
 the $ U(1)_{soc} $ symmetry with the conserved quantity $ Q_c= \sum \tilde{S}_i^y $ explicit, but at the expense of
   reaching the translational symmetry broken Hamiltonian with the $ (\pi, 0 ) $ sublattice structure in Fig.\ref{u1}b1.
    The  $ Y-y $ ground state in the original basis in Fig.\ref{u1}a becomes the $ Y-(\pi,\pi) $ Neel state
    in the $ U(1) $ basis in Fig.\ref{u1}b2. Because of the  in-compatibility of the two sublattice structures in Fig.\ref{u1}b,
    one need to introduce four HP bosons $ a,b,c,d $ corresponding to the 4-sublattice structure $ A, B, C, D $ shown in Fig.\ref{u1}b2
    respectively to perform the SWE.

\begin{figure}[!htb]
\includegraphics[width=7.5cm]{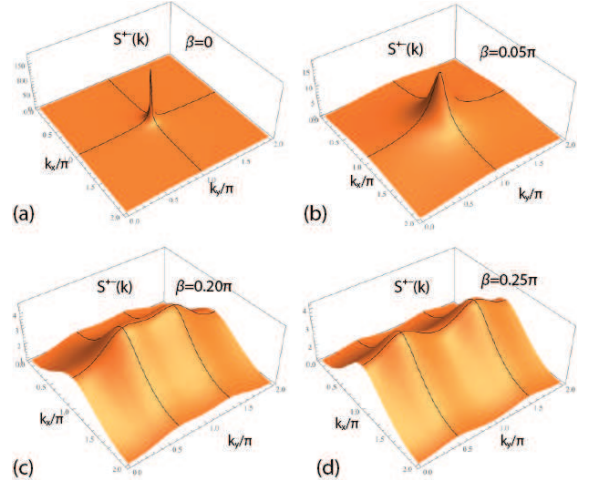}
\caption{ Transverse structure factor $ S^{+-}(\mathbf{k})$ at $ T=0 $ in the $ U(1) $ basis at the full BZ.
When $ \beta=0 $ (Abelian point), $ 0.05 \pi $ ( $ C-C_0 $ magnons ), there is a central peak at $ (\pi,\pi) $ in a Lorentzian form.
When $ \beta= 0.20 \pi, \pi/4 $  ( $ C-IC $ magnons ), the central Lorentzian peak splits into two Lorentzian ones
peaked around $ (\pi, \pi \pm k^{0}_y ) $ whose fine structures are shown in Fig.\ref{Scross}.
The longitudinal structure factor always shows a sharp peak at the ordering wavevector $ (\pi,\pi) $  in Fig.\ref{u1}b2. }
\label{Sebz}
\end{figure}

   Several physical quantities such as the magnitude of the magnetization $ M_Q(T) $, specific heat $ C_m $, the
   gaps $ \Delta $ and density of state (DOS) are gauge invariant, so are the same in both basis.
   The $ (\pi,0) $ and  $ (0,\pi) $ susceptibilities become the uniform and the $ (\pi,\pi) $ staggered
   susceptibilities respectively in the $ U(1) $ basis. The Wilson ratio is also gauge invariant after using the uniform susceptibility in the $ U(1) $ basis.
   However, the spin-spin correlations functions are gauge  dependent \cite{tqpt}.
   As shown in this section,  it is the spin-spin correlations in the $ U(1) $ basis which can map out
   the $ C-C_0 $ and $ C-IC $  relativistic magnons most efficiently \cite{rh}.
   Due to the explicit $ U(1) $ symmetries in the $ U(1) $ basis,
   the anomalous structure factors $ S^{++}=S^{--}=0$, so one need only evaluate the normal structure
    factors $ S^{+-} $. However, due to the 4-sublattice structure in Fig.\ref{u1}b2, one need to evaluate it at 4 different orbital orders at
    $Q_u=(0,0), Q_x=(\pi,0), Q_y=(0,\pi), Q_s=(\pi,\pi)$. Due to the exact relations among them in the
    Reduced Brillioun Zone (RBZ):
\begin{equation}
	S_u^{+-}(k)=S_{s}^{+-}(k+Q_s)=S_{Q_x}^{+-}(k+Q_x)=S_{Q_y}^{+-}(k+Q_y)
\label{four}
\end{equation}
    one can combine them into a single structure factor in the EBZ: $ 0 < k_x, k_y < 2 \pi $
\begin{equation}
	S^{+-}_{EBZ}(k)=\sum_{s=\pm} \frac{[1+(-1)^s\sin\theta_k](1-\gamma_k^s)}{\omega_k^s}
\label{Sfactor}
\end{equation}
   where the denominator is precisely the relativistic magnons spectrum $ \omega_k^s $ listed below Eqn.\ref{LSW},
   the numerator contains $ \gamma_k^s $ listed below Eqn.\ref{LSW} and the $ \sin\theta_k $
   is given in the unitary transformation Eqn.A2 in the appendix A.

\begin{figure}[!htb]
\includegraphics[width=7cm]{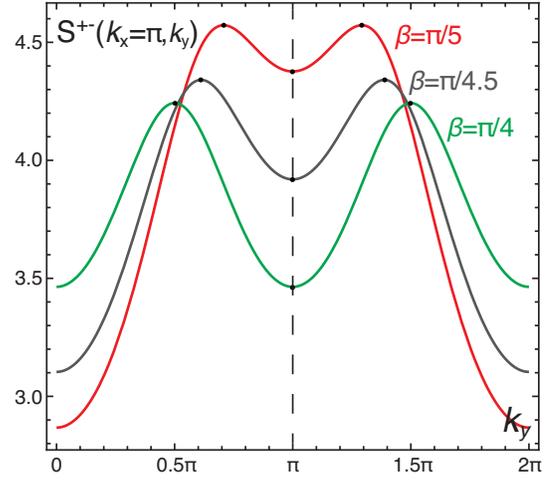}
\caption{ The cross section  $ ( k_x=\pi, k_y) $ of the transverse structure factor $ S^{+-}(\mathbf{k})$ of the C-IC near
$ \beta=\pi/4 $ in Fig.\ref{Sebz}.
At $ \beta=\pi/4 $, there are two well separated Lorentzian peaks exactly at $ (\pi, \pi \pm \pi/2) $.
When $ \beta < \pi/4 $, the two  Lorentzian peaks move closer to each other.
When $ \beta > \pi/4 $, they move apart as dictated by the Mirror symmetry ( not shown for the clarity reasons ).}
\label{Scross}
\end{figure}

 The transverse structure factor Eqn.\ref{Sfactor} at several typical $ \beta $ is shown in Fig.\ref{Sebz} and its fine structure near $ \beta=\pi/4 $ is shown
 in Fig.\ref{Scross}. When $ 0< \beta < \beta_1 $ or $ \beta_2 < \beta < \pi/2 $, the $ C-C_0 $
 leads to a central peak at $ \mathbf{k}=(\pi,\pi) $.
  At the Abelian point $ \beta=0 $ in Fig.\ref{Sebz}a, $ S^{+-}(q) \sim \frac{\sqrt{2}}{q} $ where
  $ k=(\pi,\pi) + q $. Obviously, the singularity  at $ (\pi,\pi) $ is due to the infra-red divergence of the Goldstone mode
  $ \omega=ck $ in Eqn.\ref{Sfactor}. At a small $ \beta $ in the $ C-C_0$ regime in Fig.\ref{Sebz}b:
\begin{equation}
   S^{+-}(q)_{C-C_0} \sim \frac{1}{ \sqrt{ v^{2}_x q^2_x+  v^{2}_y q^2_y + \Delta^2(\beta)} }
\end{equation}
  where the $ \Delta(\beta) \sim \sqrt{2} \beta $ is the gap opening due to the small $ \beta $ listed in Eq.\ref{disper2} and
  shown in Fig.\ref{groundex}b.

  In Fig.\ref{Sebz}c, when $ \beta_1 < \beta < \beta_2 $,
  the  C-IC starts to splits the central peak into two peaks located around
  its two minima $ \mathbf{k}=(\pi, \pi \pm k^{0}_y ) $ shown in Fig.\ref{groundex}a,
  the $ \mathbf{k}=(\pi,\pi) $ becomes a saddle point being maximum along the $ k_x $ direction, minimum along the $ k_y $ direction.
  In Fig.\ref{Sebz}d, at $ \beta=\pi/4 $, the two peaks are exactly located at the two minima $ ( \pi, \pi \pm \frac{\pi}{2} ) $ of the $C-IC$
  shown in Fig.\ref{groundex}a, each of the two well separated Lorentzian peaks
  is given by:
\begin{equation}
   S^{+-}_{C-IC}(q) \sim \frac{3/2}{ \sqrt{ ( q^2_x+q^2_y)/4 + 1/2 } }
\end{equation}
  where $  \Delta(\beta=\pi/4)= 1/2 $ is the largest gap at $ \beta=\pi/4 $ shown in Fig.\ref{groundex}b.
  As shown in Fig.\ref{Scross}, the two Lorentzian peaks are moving closer when $ \beta < \pi/4 $  or apart when $ \beta > \pi/4 $.
  So in the C-IC regime, the structure factor maps out precisely the dispersions of the  C-IC relativistic magnons which are
  completely due to quantum fluctuations and
  intrinsically embedded in the quantum Y-y ground state at $ T=0 $.

  It is constructive to contrast to RFHM  where the sublattice structure of the transformed Hamiltonian $ (\pi, 0 ) $ is
  compatible with the classical FM state in the $ U(1) $ basis \cite{rh}, so one need only introduce two HP bosons to perform SWE.
  So one only need to form a uniform $ S^{+-}_{u}(k) $ and a $ (\pi,0) $  staggered  transverse structure factor  $ S^{+-}_{s}( k ) $
  in the RBZ listed in Eqn.31 in \cite{rh}.
  One can also establish the exact relation $ S^{+-}_{u}(k)= S^{+-}_{s}( k + (\pi,0) ) $, so one can combine them into a
  single structure factor in the EBZ. It is a Gaussian exponentially suppressed by $ e^{-\Delta/T} $,
  peaked at $ (0, \pm k^{0}_y ) $  with a temperature dependent width
  $ \sigma_x= \sqrt{ m_{x}(\beta) T } $ due to the thermal fluctuations at a finite $ T $.
  This is  because there is no quantum fluctuations at $ T=0 $. The $ Y-x $ state is an exact eigenstate.
  The C-IC magnons do NOT exist at $ T=0 $, so they need to be thermally excited, so can only be detected at a finite $ T $.

\section{ Weak coupling Y-y state, low energy fermionic excitations and weak to strong crossover }

\begin{figure}
\includegraphics[width=0.55\linewidth]{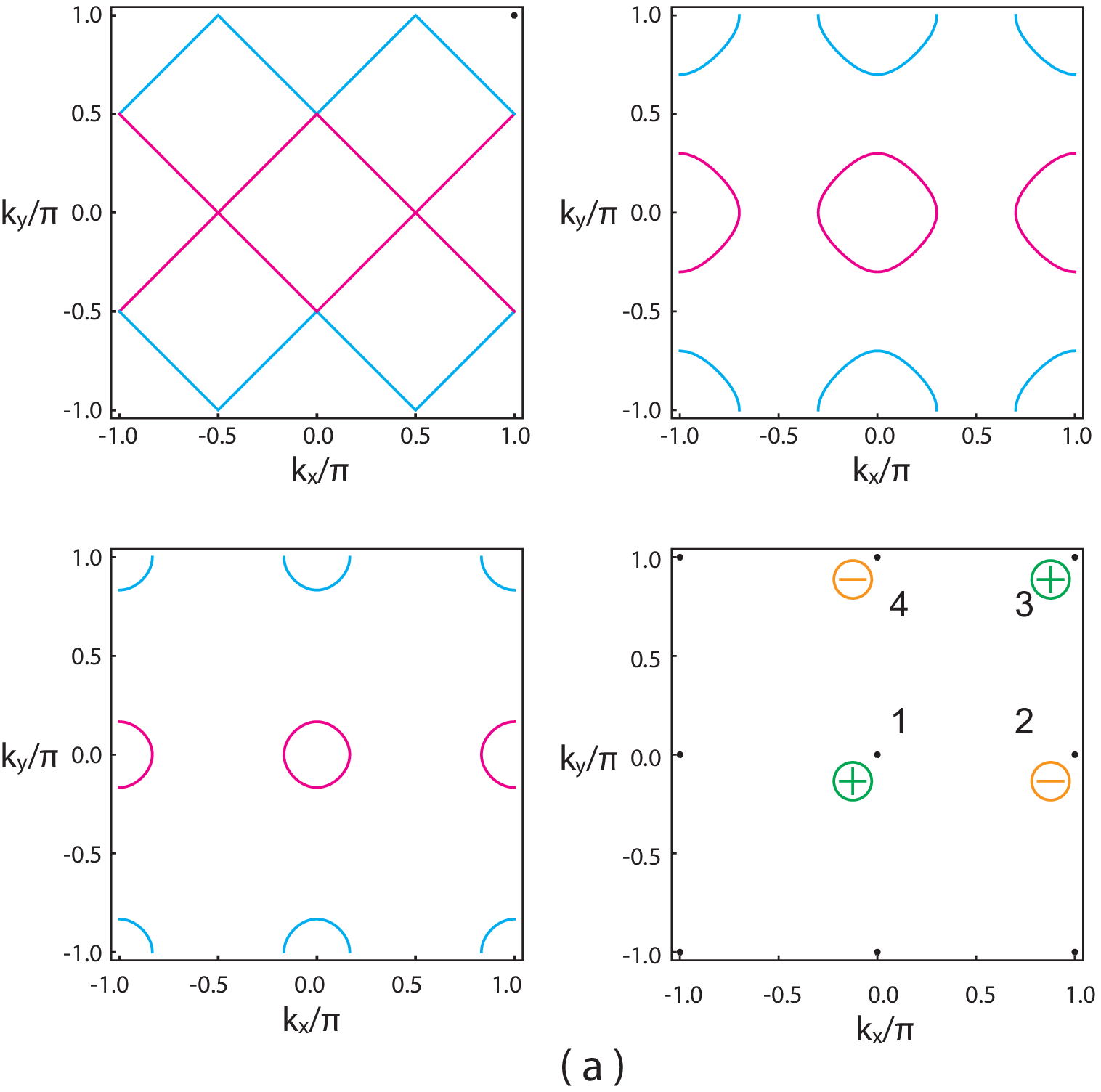}
\hspace{0.2cm}
\includegraphics[width=0.40\linewidth]{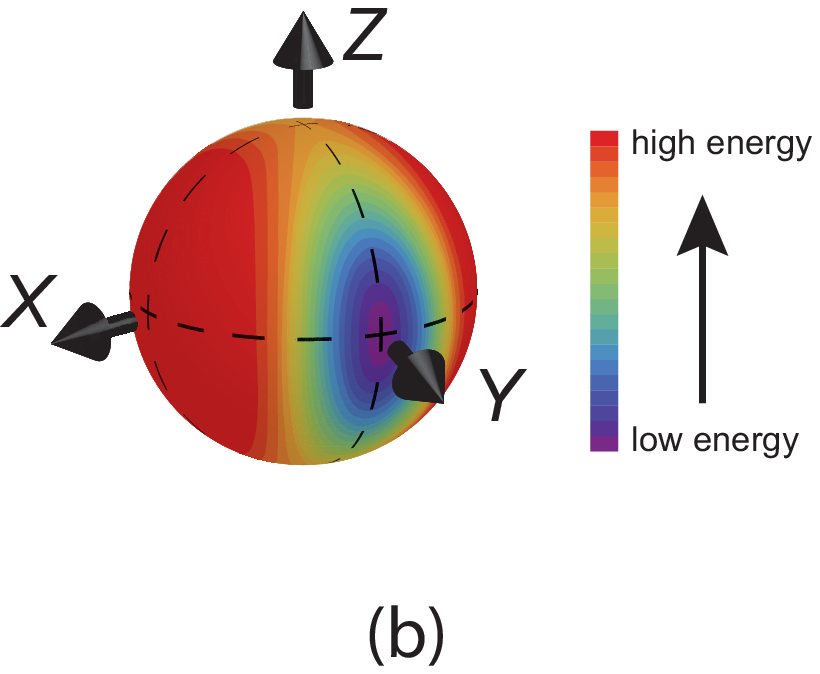}
\caption{
(a) The Fermi surface (FS) evolves along the line
$(\alpha=\pi/2,\beta)$ at $\beta=0,\pi/5,2\pi/5,\pi/2$.
At $\beta=\pi/2$, there are 4 Dirac fermions located at
$\mathbf{K}_1=(0,0), \mathbf{K}_2=(\pi,0), \mathbf{K}_3=(\pi, \pi)$ and  $\mathbf{K}_4=(\pi,0)$
labeled as $1,2,3,4 $ with the topological winding numbers $\pm1$.
There are FS nesting away from  $\beta=\pi/2$.
Purple (green) is particle (hole) surface.
(b) The ground-state energy as a function of magnetic orientation
$\mathbf{M}=(X,Y,Z)$ at the orbital order $\mathbf{Q}=(0,\pi)$ with the parameter $U=0.2t, \beta=\pi/6$.
The position on the sphere indicates the Spin orientation.
Red ( purple ) color means higher ( lower ) energy. The figure shows that
the $ Y $ spin-orientation is the ground state at the orbital order $\mathbf{Q}=(0,\pi)$.
In fact, the $ Y-y $ state is the global ground state when considering all the other possible orbital orderings. }
\label{squaredash}
\end{figure}

So far, we focused on the strong coupling expansion at $U\gg t$ where the RAFHM Eqn.\ref{rh} holds and the charge degree of freedoms are frozen.
It is also important to start from the weak coupling limit﹛$ U \ll t $ where one need to also consider charge fluctuations
and study how it approaches the strong coupling limit.
Using the identity $n_{i\uparrow}n_{i\downarrow}=\frac{1}{2}n_i-\frac{2}{3}\mathbf{S}_{i}^2$
to  explicitly keep the spin SU(2) symmetry of the Hubbard interaction in Eqn.\ref{hubbardint},
one can introduce a magnetic order parameter $\mathbf{M}_{i} $ to decouple the interaction term:
\begin{equation}
	\mathcal{H}_{\mathbf{M}}
	=-t\!\sum_{\langle i,j\rangle}
		(c_{i\sigma}^\dagger
		U_{ij}^{\sigma\sigma'}\!\!
		c_{j\sigma'}+h.c.)
	+\frac{3}{8U} \sum_{i}\mathbf{M}^{2}_{i}
	+\sum_{i} \mathbf{M}_{i} \cdot \mathbf{S}_{i}
\label{hubbardmag}
\end{equation}
The evolution of the non-interacting Fermi surfaces (FS) along the line $ (\alpha=\pi/2, \beta ) $ is shown in Fig.\ref{squaredash}a.
Due to the FS nesting conditions at the half-filling shown in Fig.\ref{squaredash},
{\sl any weak interaction} will open gaps to the non-interacting FS when $ \beta \neq \pi/2 $.
So one can perform a well controlled weak coupling analysis to determine the spin-orbit orders of the ground state
and also the excitation spectra.

 From the FS geometry in Fig.\ref{squaredash}a, there can only be four possible orbital orders
 $\mathbf{Q}_1=(0,0), \mathbf{Q}_2=(\pi,0), \mathbf{Q}_3=(0,\pi), \mathbf{Q}_4=(\pi,\pi)$.
 Substituting the order parameter $ \mathbf{M}_i=\mathbf{M}e^{i\mathbf{Q}\cdot\mathbf{r}_i}$
 where $\mathbf{Q}=\mathbf{Q}_i, i=1,2,3,4 $ and  $\mathbf{M}=(X,Y,Z)$ into Eqn.\ref{hubbardmag}
 leads to the mean field Hamiltonian: $ H_{MF}= H_0+ \frac{3N|\mathbf{M}|^2}{8U}$ and
\begin{align}
	H_{0}=\frac{1}{2}
	\sum_k
	\begin{pmatrix}
		c_{k}^\dagger &c_{k+Q}^\dagger
	\end{pmatrix}
	\begin{pmatrix}
		T_k &M^a\sigma^a\\
		M^a\sigma^a &T_{k+Q}\\
	\end{pmatrix}
	\begin{pmatrix}
		c_{k}\\
		c_{k+Q}\\
	\end{pmatrix}
\label{matrix}
\end{align}
    where $ T_k=-4t[\cos\beta\cos k_y\!-\!\sigma^x\sin k_x\!-\!\sigma^y\sin\beta\sin k_y] $ is the kinetic part of $ H_0 $
    encoding the SOC parameters $ (\alpha=\pi/2, \beta ) $.

For $\mathbf{Q}_3=(0,\pi)$, diagonalizing the $ 4 \times 4 $ matrix in Eqn.\ref{matrix} leads to 4 fermionic energy levels
$ \pm \epsilon_1, \pm \epsilon_2 $.
It is an insulating state with the P-H symmetry.
Due to the lack of the spin $ SU(2) $ symmetry, the minimization procedures are much more involved than those with the symmetry.
We also take the "divide and conquer " strategy: first fixing the spin orientation and find the optimal magnitude and energy in the subspace,
then determine the optimal orientation and magnitude. The results for the general spin orientation $\mathbf{M}=(X,Y,Z)$ is
shown in Fig.\ref{squaredash}b.
The global ground state has the $ (0,Y,0) $ spin orientation, it is nothing but the $ Y-y $ state which respects the $ U(1)_{soc} $ symmetry.
It also supports 2 branches gapped fermionic excitations listed in Eq.\ref{fer}.
By repeating the calculations for $ \mathbf{Q}_4=(\pi,\pi) $, we find the lowest spin-orientation is $ X-(\pi,\pi) $
which breaks the $ U(1)_{soc} $ symmetry. By applying the $ U(1)_{soc} $ symmetry operator, one can see $ X-(\pi,\pi) $ is degenerate with
$ Z-(0,\pi) $ which, of course, has higher energy than the $ U(1)_{soc} $ symmetric $ Y-y $ state as shown in Fig.\ref{squaredash}b.

For $ \mathbf{Q}_2=(\pi,0) $ or $ (0,0) $, it is the magnetic ordering in the P-P channel
or hole-hole ( H-H ) channel, so breaks the P-H symmetry, both need a finite $ U_c >0 $ to reach a metallic state
with only partial fillings of
all the 4 fermionic bands $ \epsilon_{i}, i=1,2,3,4 $. It
has much higher energies than those insulating states in the P-H channel.
So we conclude that the $ Y-y $ states is indeed the global ground state at the weak coupling.

\begin{figure}[!htb]
\includegraphics[width=7cm]{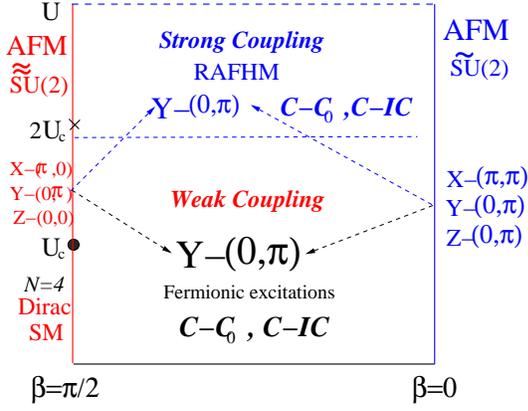}
\caption{  At any weak $U>0$, the Y-y state emerges as the ground state at any Non-Abelian point.
There are both low energy fermionic excitations and $ C-C_{0}, C-IC $ relativistic magnons
excitations. However, there are no Mirror symmetry anymore in the weak Y-y state.
There is only a crossover  from the weak to the strong coupling.}
\label{weak}
\end{figure}

  As shown in Fig.\ref{weak}, due to the $\tilde{SU}(2) $ and $\tilde{\tilde{SU}}(2) $ symmetry
  at the two Abelian points $ \beta=0, \pi/2 $ respectively, the Y-y state is degenerate with the other two states.
  However, away from them, the FS nesting conditions in Fig.\ref{squaredash} at the half-filling favors the Y-y state
  which also supports the low energy fermionic excitations in Eq.\ref{fer}.
  So the specific heat in Eqn.\ref{CmT} will also receive the contributions from the fermionic part.


Following the procedures in \cite{ssdw}, splitting the magnetic fluctuations into one longitudinal and two transverse components and
performing  Gaussian fluctuations above the $ Y-y $ state, we can also identify the $ C-C_0 $ and $ C-IC $ magnons
from the poles of the dynamic transverse spin structure factor $ S^{+-}(\vec{k}, \omega ) $.
Of course, at the two Abelian points, the $ C-C_0 $ reduce to the two Gapless goldstone modes.
They should smoothly crossover to those in Fig.\ref{groundex} achieved by the SWE in the strong coupling regime shown in Fig.\ref{squaredash}b.
Note that the weak $ Y-y $ state still respects the spin-orbital coupled $ U(1)_{soc} $ symmetry $[H_f,\sum_i(-1)^{i_x}c_i^\dagger\sigma^y c_i]=0 $.
However, the Mirror symmetry valid in the strong coupling limit does not hold anymore in the weak coupling limit.
So there is a crossover from the weak to strong coupling where all the physical quantities evolve from having $ {\cal M} $ asymmetry to
owning $ {\cal M} $ symmetry with respect to $ \beta=\pi/4 $.
The next order terms $ \sim t^4/U^3 $ in the strong coupling expansion which include a ring exchange term
around a fundamental square do not have such a Mirror symmetry.  They may be needed to describe the crossover in Fig.\ref{weak}.
The crossover driven by $ U>0 $ is dual to the BCS to BEC crossover driven by $ U<0 $ in SOC coupled Fermi systems discussed
in \cite{pairing}.


\section{  High temperature expansions, electronic and spin  Wilson loops at weak and strong coupling. }
 In the classical fluctuation regime shown in the Fig.\ref{finiteT},  the $ Y-y $ state to the paramagnet transition in the RAFHM
 is in the same universality class as that from $ Y-x $ state to the paramagnet transition in the RFHM in \cite{rh} which was shown to be in the 2d
 Ising universality class \cite{rhht}.
 In the high temperature regime in the Fig.\ref{finiteT},
 from the high temperature expansions in the $ T_c \sim J \ll T \ll U $ limit,
 one can easily establish the relation between the free energy of RAFHM  and that of the RFHM in \cite{rh}:
\begin{equation}
 F^{H}_{\rm RAFHM}[J]= F^{H}_{\rm RFHM}[-J]
\end{equation}
 Of course, the above relation breaks down in the symmetry breaking low temperature phases.
 Taking Eqn.18 in \cite{rh} and changing $ J $ to $ -J $ leads to high temperature expansion
 of the RAFHM:
\begin{equation}
	C_m/N=\frac{3}{8} (\frac{J}{T})^2+\frac{3}{16} ( \frac{J}{T} )^3
	+\frac{6 W_R-39}{128} ( \frac{J}{T} )^4
\label{hightheat1}
\end{equation}
 where $ W_R=2 \cos4 \beta +1 $ is the Wilson loop around the fundamental square given in the Sec.II.
 The discussions below Eqn.18 in \cite{rh} also apply here.

In the weak coupling $U\ll t \ll T $ limit, following the method in \cite{ahe},
we perform the high temperature expansion in the limit $  T \gg t $ directly on the fermionic model Eq.\ref{hubbardint}
to evaluate its specific heat:
\begin{align}
	C_f(T)/N=\frac{4t^2}{T^2}-(16+2W_f )\frac{t^4}{T^4}+\cdots
\label{hightheat2}
\end{align}
  which establishes its connection with the electronic Wilson loops $ W_f $ given in Sec.I.
  Note that Eq.\ref{hubbardint} is invariant under $t\to-t$, so there is no odd power of $t/T$ term in the expansion,
  in contrast to Eqn.\ref{hightheat1} which has odd power of terms.
The term in the $ (t/T)^4$ power proportional to the electronic Wilson loop $ W_f $ comes from the fermion hopping around a closed plaquette
in the square lattice.
Because $ U\ll t$, so the interaction effects may be dropped in Eqn.\ref{hightheat2}, so it is essentially a free fermions hopping in a non-abelain gauge
potentials. So the crossover driven by the interaction $ U $ at the low temperature $ Y-y $ state in Fig.\ref{squaredash}
can also be partially seen by looking at the specific heat crossover from Eqn.\ref{hightheat2} to Eqn.\ref{hightheat1}
in the  high temperature paramagnet state.

\section{ Experimental Realizations and Detections: }

  In condensed matter systems, as said in the introduction, any of the linear superpositions
  of the Rashba SOC $ k_x \sigma_x + k_y \sigma_y $ and Dresselhaus SOC $ k_x \sigma_x - k_y \sigma_y $ always exists
  in various noncentrosymmetric 2d or layered materials. In momentum space, such a linear combination
  $ \alpha k_x \sigma_x + \beta k_y \sigma_y $ can be written as the the kinetic term
  in Eq.\ref{hubbardint} in a periodic array of adsorbed ions with the SOC parameter $ (\alpha,\beta ) $
  where the  anisotropy can be adjusted by the strains, the shape of the surface or gate electric fields.
  The interaction strength $ U $ in Eq.\ref{hubbardint} ranges from weak to strong in different materials \cite{sherev,aherev}.
  So all the phenomena in Fig.\ref{weak} can be observed in these materials.

  In the cold atom systems, in view of recent experimental advances to realize 2d Rashba or Dresselhaus SOC \cite{expk40,2dsocbec},
  both the original and the $ U(1) $ basis can be realized. Both gauge-invariant and non gauge-invariant
  quantities can be measured \cite{tqpt}. The gauge invariant quantities such as  specific heat $ C_m $ \cite{heat1,heat2}, the
  gaps $ \Delta $ and  the DOS \cite{dosexp}, the Wilson ratio can be detected by the corresponding experimental tools.
  The magnetization $ M_Q(T) $, the $ (\pi,0) $ and  $ (0,\pi) $ susceptibilities  can be detected by the longitudinal atom
  or light Bragg spectroscopies \cite{lightatom1,lightatom2}.
   In the $ U(1) $ basis in Fig.\ref{u1}b, one need to measure the transverse structure factor at the four different ordering wavevectors
  $ Q_u=(0,0), Q_x=(\pi,0), Q_y=(0,\pi), Q_s=(\pi,\pi)$ in Eq.\ref{four} by the transverse atom  or
  light Bragg spectroscopies \cite{lightatom1,lightatom2}  to get
  the whole transverse structure Eqn.\ref{Sfactor} in the EBZ. Before reaching $ T < T_c \sim J $, the specific heat measurement
 \cite{heat1,heat2} at high temperatures to
  determine the whole sets of fermionic or magnetic Wilson loops order by order in $ t/T $ Eqn.\ref{hightheat2}
  or $ J/T $ Eqn.\ref{hightheat1} could be performed easily.
  However, so far, the interaction in these cold atom experiments is still in weak coupling limit.
  So the weak Y-y state, both its fermionic and magnon excitations in Fig.\ref{weak} can still be observed by various detection methods
  \cite{jin,heat1,heat2,lightatom1,lightatom2} in the current available weak coupling limit.
  Because there is only a crossover from the weak to the strong coupling, the results on magnons achieved in the strong coupling limit
  still hold qualitatively in the weak coupling limit.  When the heating issue
  is completely overcame as the interaction strength is tuned to the strong coupling limit, the RAFHM Eqn.\ref{rh} can be realized
  and all the strong coupling results achieved here can be detected quantitatively.


\section{ Discussions and conclusions:}

     There are previous theoretical works to study strongly correlated spinless bosons in Abelian gauge fields \cite{bosonmsg,honey1,dual1,dual2,dual3}
 and spinor bosons in non-Abelian gauge fields \cite{cwu,classdm1,classdm2,rh,rhh,rhht}.
 The topological quantum phase transitions of non-interacting fermions driven by a Rashba type of SOC are investigated
 in a honeycomb lattice \cite{tqpt}. Various itinerant phases and phase transitions
 of repulsively interacting fermions subject to Weyl type of SOC in a 3d continuum were studied in \cite{ssdw}.
 The BCS to BEC crossover of attractively interacting fermions tuned by the strengths of various forms of SOC in 2d and 3d continuum were
 explored in \cite{pairing}. However, so far, there are very few works to study the possible dramatic effects of SOC on strongly
 correlated electron systems on lattice systems. In this paper,  we investigate the system of interacting fermions
 at half filling hopping in a 2 dimensional square lattice subject to any combinations of Rashba and Dresselhaus SOC described by Eq.\ref{hubbardint}.
 In the strong coupling limit, we reach a novel quantum spin model named Rotated Anti-Ferromagnetic Heisenberg
 model (RAFHM) Eq.\ref{rh} which is a new class of quantum spin models. Along the
 anisotropic line $ (\alpha=\pi/2, \beta ) $, its ground state is a new kind of spin and bond correlated
 magnetic state called Y-y state in Fig.\ref{u1}a
 which supports a novel excitation called C-IC magnons in a large SOC parameter regime $ \beta_1 < \beta < \beta_2 $ in Fig.\ref{groundex}a.

     The C-IC magnons in the RAFHM stand for the short-ranged IC seeds embedded in a commensurate long-range ordered Y-y state.
     Their parameters such as the minimum positions $ ( 0, k^{0}_y ) $, gap, velocities $ v_x, v_y $ can be precisely
     measured by the peak positions, the width and Lorentzian shape of the transverse structure factor at $ T=0 $ respectively.
     In this sense, they resemble quite closely to an elementary particle resonance in scattering cross sections in particle physics.
     It remains interesting to see how these seeds response under various external probes.
     To transfer the short-ranged IC order to a long-ranged one, one need to apply an external probe to drag it out and then drive its condensation.
     We will study how these magnons response under a finite $ ( \pi,0) $  longitudinal field $ h_y $ which
     couples to the conserved quantity  and still keeps the spin-orbital coupled $ U(1) $ symmetry or two different $ ( \pi,0) $  transverse fields $ h_x $ and $ h_z $ which breaks it explicitly.

      It may be necessary to point out the RAFHM Eq.\ref{rh} is explicitly for  spin $ s=1/2 $.
      However, the RFHM in Refs.\cite{rh,rhh,rhht,rhrashba} is for any spin $ s=N/2 $. As argued in \cite{rhh},
      the critical temperature $ T_c/J \sim 2S $, so increasing the spin is an very effective way to raise the critical temperatures.
      It is known that if putting $ s=3/2 $ fermions on a lattice without SOC, the resulting spin model in the strong coupling
      limit at half filling \cite{so5} has a higher symmetry such as $ SO(5) $ instead of $ SU(2) $, it has even large quantum fluctuations due to the enlarged symmetry.  It remains important to achieve any spin $ S $ RAFHM.

      As mentioned in the introduction, starting from the results achieved in this paper along the solvable
      line $ (\alpha=\pi/2, \beta) $, we will investigate the quantum or topological phenomena at
      a generic equivalent class  $ ( \alpha, \beta ) $ including the isotropic Rashba or  Dresselhauss  lines $ \alpha=\pm \beta $
      at both the weak and strong coupling limit.
      Recently, the same "divide and conquer " approach has been employed to map out the very rich and novel phenomena of RFHM
      in the generic SOC parameter $ ( \alpha, \beta ) $  in \cite{rhrashba}.
      As shown in this paper, the RAFHM display quite different phenomena than those
      in the RFHM \cite{rh} along the solvable line $ ( \alpha=\pi/2, \beta) $. So
      we expect that the global phase diagram of RAFHM Eq.\ref{rh}  may also show quite different phenomena than those of RFHM in \cite{rhrashba}.
      Expansion to the $ t^3/U^2 $ order which includes  ring exchange terms around a square plaquette may also be necessary to
      study the possible phases and phase transitions from weak to strong coupling limit.
      The SOC could provide a new mechanism to lead to Spin liquid phases with topological orders even in a bipartite lattice.
      Possible topological spin liquid phases in a honeycomb lattice with three SOC parameters $ ( \alpha, \beta, \gamma) $
      need to be explored.

{\bf Acknowledgement }

We acknowledge  AFOSR FA9550-16-1-0412 for supports. The work at KITP was supported by NSF PHY11-25915.
W.M. Liu is supported by NSFC under Grants No. 10934010 and No. 60978019, the NKBRSFC under Grants No. 2012CB821300.

\appendix

\begin{widetext}

In this appendix, we provide some technical details on the results achieved in the main text:
(1) The symmetry and symmetry breaking analysis of the fermionic and the RAFHM, followed by a specific linear spin wave expansion (LSWE) at $1/S $ order.
(2) The $1/S $ correction to the LSWE results.
(3) The structure of the quantum Y-y ground state which encodes the C-IC magnons.
(4) The fermionic excitations in  the Y-y state at weak coupling $ U \ll t $.

\section{ Symmetry analysis and  the Linear Spin Wave  Expansion }

The general approach to investigate an interesting model is to first present analysis an exact  symmetry analysis which will lead to
some non-trivial exact results which will put constraints on any specific calculations \cite{ssdw}. Here, the specific calculation is the systematic
spin wave calculations in terms of $ 1/S $.

At a generic $ (\alpha, \beta ) $, the fermionic model Eq.\ref{hubbardint} has the Time reversal symmetry $ {\cal T} $:
$ \vec{k} \rightarrow -\vec{k}, \vec{S} \rightarrow -\vec{S} $, translational symmetry and three spin-orbital coupled $ Z_2 $ symmetries:
    (1) $ {\cal P}_x $ symmetry: $ S^{x} \rightarrow S^{x}, k_y \rightarrow - k_y, S^{y} \rightarrow - S^{y},  S^{z} \rightarrow - S^{z} $.
    (2) $ {\cal P}_y $ symmetry: $ S^{y} \rightarrow S^{y}, k_x \rightarrow - k_x, S^{x} \rightarrow - S^{x},  S^{z} \rightarrow - S^{z} $.
    (3) $ {\cal P}_z $ symmetry: $ k_x \rightarrow - k_x, S^{x} \rightarrow - S^{x}, k_y \rightarrow - k_y, S^{y} \rightarrow - S^{y},
    S^{z} \rightarrow S^{z} $ which is also equivalent to a joint $ \pi $ rotation of both the spin and the orbital  around $ \hat{z} $ axis.
At the  Rashba or Dresselhaus point $ \alpha= \pm \beta $, the $ {\cal P}_z $ symmetry is enlarged to the
spin-orbital coupled symmetry $ C_4 \times C_4 $ symmetry which is a joint $ \pi/2 $ rotation of both the spin and the orbital  around $ \hat{z} $ axis.
Along the line $(\alpha=\pi/2, \beta)$, there is also an enlarged spin-orbital coupled $ U(1)_{soc} $ symmetry $[H_f,\sum_i(-1)^{i_x}c_i^\dagger\sigma^y c_i]=0 $.
Of course, at the two Abelian points, the $ U(1)_{soc} $ symmetry is enlarged to the $ SU(2) $ symmetry in the corresponding rotated basis.

The gauge invariant fermionic Wilson loop around an elementary square is the same as the bosonic case \cite{rh}
$W_f=2-4\sin^2\alpha\sin^2\beta$ which stands for the non-Abelian flux through the square.
The $R$-matrix Wilson loop $W_R $ around a fundamental square is defined as
$W_R={\rm Tr}[R_xR_yR_x^{-1}R_y^{-1}]=(W_f)^2-1$ which can be used
to characterize the equivalent class and frustrations in the RAFHM Eq.\ref{rh}
The $W_R=3$ ($W_R<3$) stands for the Abelian (non-Abelian) points.
The relations between two sets of Wilson loops are
in two-to-one relation due to the coset $ SU(2)/Z_2=SO(3) $.
As expected, the RAFHM with $ J >0 $ should display dramatically different physics
than the RFHM model with $ J < 0 $ studied in Ref.\cite{rh}.

 Now we get to the specific spin wave calculations which can be contrasted to the above exact statements.
 After introducing two HP bosons $ a,b $ corresponding to the two sublattices $ A, B $ in Fig.\ref{u1}a,
 we obtain the Hamiltonian at the LSW $ 1/S $ order:
\begin{equation}
    \mathcal{H}_2  =  2\sum_{\mathbf{k}}
    (a_\mathbf{k}^\dagger a_\mathbf{k}
    +b_\mathbf{k}^\dagger b_\mathbf{k})
    +\frac{1}{2}\sum_{\mathbf{k}}[
    \cos k_x(a_\mathbf{-k}a_\mathbf{k}
	    +b_\mathbf{-k}b_\mathbf{k})
    +\cos(k_y+2\beta)a_\mathbf{-k}b_\mathbf{k}
    +\cos(k_y-2\beta)b_\mathbf{-k}a_\mathbf{k}
    +h.c.] ~~~~~~~
\label{h2m}
\end{equation}

    We first perform a unitary transformation
\begin{align}
    \begin{pmatrix}
	a_\mathbf{k}\\
	b_\mathbf{k}\\
    \end{pmatrix}
    =U_\mathbf{k}
    \begin{pmatrix}
	\bar{a}_\mathbf{k}\\
	\bar{b}_\mathbf{k}\\
    \end{pmatrix},
    \quad
    U_\mathbf{k}=
    \begin{pmatrix}
	\sin\frac{\theta_\mathbf{k}}{2}
	 &\cos\frac{\theta_\mathbf{k}}{2}\\
	-\cos\frac{\theta_\mathbf{k}}{2}
	 &\sin\frac{\theta_\mathbf{k}}{2}\\
	\end{pmatrix}
\label{unit}
\end{align}
where $
	\sin\theta_k=\frac{\cos k_x}{\sqrt{\cos^2 k_x+\sin^22\beta\sin^2k_y}},
	\quad
	\cos\theta_k=\frac{\sin2\beta\sin k_y}{\sqrt{\cos^2 k_x+\sin^22\beta\sin^2k_y}} $ is identical to that used in the RFHM \cite{rh} to
    cast the Hamiltonian Eqn.\ref{h2m} into a simple form:
\begin{equation}
    \mathcal{H}_2 = \sum_k
    \begin{pmatrix}
	\bar{a}_k^\dagger\\
	\bar{a}_{-k}\\
    \end{pmatrix}^\intercal\!
    \begin{pmatrix}
	1 &\lambda_k^-\\
	\lambda_k^- &1\\
    \end{pmatrix}
    \begin{pmatrix}
	\bar{a}_k\\
	\bar{a}_{-k}^\dagger\\
    \end{pmatrix}
    +\sum_k
    \begin{pmatrix}
	\bar{b}_k^\dagger\\
	\bar{b}_{-k}\\
    \end{pmatrix}^\intercal\!
    \begin{pmatrix}
	1 &\lambda_k^+\\
	\lambda_k^+ &1\\
    \end{pmatrix}
    \begin{pmatrix}
	\bar{b}_k\\
	\bar{b}_{-k}^\dagger\\
    \end{pmatrix}
\label{h2mu}
\end{equation}
where $   \lambda_k^{\pm}=\pm{\rm sgn}(\cos k_x)\gamma_k^{\pm} $ and
the $ \gamma_k^{\pm}=\frac{1}{2} [\cos2\beta\cos k_y
\pm\sqrt{\cos^2 k_x+\sin^22\beta\sin^2k_y} ]$ which is also listed below Eq.\ref{LSW}.

   Then we perform the Bogoliubov transformations
\begin{equation}
    \begin{pmatrix}
	\bar{a}_k\\
	\bar{a}_{-k}^\dagger\\
    \end{pmatrix}
    =
    \begin{pmatrix}
	\cosh\phi_k^- &\sinh\phi_k^-\\
	\sinh\phi_k^- &\cosh\phi_k^-\\
    \end{pmatrix}
    \begin{pmatrix}
	\alpha_k\\
	\alpha_{-k}^\dagger\\
    \end{pmatrix},
    \quad
    \begin{pmatrix}
	\bar{b}_k\\
	\bar{b}_{-k}^\dagger\\
    \end{pmatrix}
    =
    \begin{pmatrix}
	\cosh\phi_k^+ &\sinh\phi_k^+\\
	\sinh\phi_k^+ &\cosh\phi_k^+\\
    \end{pmatrix}
    \begin{pmatrix}
	\beta_k\\
	\beta_{-k}^\dagger\\
    \end{pmatrix}
\label{bog}
\end{equation}
where  $  2\phi_\mathbf{k}^-
    =-{\rm arcsinh}(\lambda_\mathbf{k}^-/\omega_\mathbf{k}^-),
    \quad
    2\phi_\mathbf{k}^+
    =-{\rm arcsinh}(\lambda_\mathbf{k}^+/\omega_\mathbf{k}^+) $
    to transform the Hamiltonian Eqn.\ref{h2mu} to the diagonal form
    Eqn.\ref{LSW}:
\begin{align}
    \mathcal{H}_2=
    \sum_\mathbf{k}
	(\omega_\mathbf{k}^-+\omega_\mathbf{k}^+-2)
    +2\sum_\mathbf{k}
	(\omega_\mathbf{k}^-\alpha_\mathbf{k}^\dagger\alpha_\mathbf{k}
	+\omega_\mathbf{k}^+\beta_\mathbf{k}^\dagger\beta_\mathbf{k})    \nonumber
\end{align}
   where $\omega_\mathbf{k}^\pm=\sqrt{1-(\gamma_\mathbf{k}^{\pm})^2}$.

   When $ \beta < \pi/4 $, $ \omega^{+}_{k} < \omega^{-}_{k} $, when $ \beta > \pi/4 $, $ \omega^{+}_{k} > \omega^{-}_{k} $,
   at $ \beta = \pi/4 $, $ \omega^{+}_{k} = \omega^{-}_{k} $. In the main text, for notational simplicity, we assume $ \omega^{-}_{k} $
   is always the lower branch. The minima of excitation spectrum $\mathbf{k}^0=(0,k_y^0)$ can be determined as:
\begin{eqnarray}
     k_y^0(\beta) = \left \{ \begin{array}{ll}
	 0,~~  0<\beta<\beta_1
     \\
	 \pm\arccos[\sqrt{1+\sin^22\beta}/\tan2\beta],~~ \beta_1<\beta<\pi/4
    \end{array}     \right.
\label{k0y}
\end{eqnarray}
    where  $k_y^0(\beta)=k_y^0(\pi/2-\beta), \pi/4< \beta <\pi/2 $ which is shown in Fig.\ref{groundex}a.
    Note that the $\beta_1$ and $\beta_2$ coincide with those in RFHM \cite{rh}.

    Expanding around the minima lead to the relativistic form Eq.\ref{disper}
    where the mass and the two velocities are given by:
\begin{align}
	&\Delta^2=1-(1+\cos2\beta)^2/4,\quad
	v_x^2=\cos^2\beta/2,\quad
	v_y^2=\cos^2\beta(\cos^22\beta+\cos2\beta-1)/2,\quad
	0<\beta<\beta_1\\
	&\Delta^2=(3-\csc^22\beta)/4,\quad
	v_x^2=1/(4\sin^22\beta),\quad
	v_y^2=(\sin^42\beta+\sin^22\beta-1)/(4\sin^4 2\beta),\quad \beta_1<\beta<\pi/4
\label{disper2}
\end{align}
    which are shown in Fig.\ref{groundex}b.
    It is easy to check that as $\beta\to0$, $ \Delta \sim \sqrt{2} \beta \rightarrow 0, v_x \rightarrow 1/\sqrt{2}, v_y \rightarrow 1/\sqrt{2} $.
    So the dispersion $\omega_q\to v|\mathbf{q}|$ as it should be.
    Note that the QC regimes in Fig.\ref{finiteT} is defined as $ \beta \ll T $.

    The dispersion relation of both $ C-C_0 $ and C-IC take the relativistic form with the mass $ \Delta $
   and two velocities $v_x$ and $v_y$.
   The anisotropy between the two velocities at $\beta\neq\pi/4$ are irrelevant under the Renormalization group (RG),
   so the relativistic invariance is restored under the RG.
   In a sharp contrast, all the magnons in the RFHM \cite{rh} are non-relativistic gapped particles
   with a gap $\Delta$ and  two effective masses $m_y \geq m_x $.
   Note that it is the Bogoliubov transformation Eqn.\ref{bog} which leads to quantum fluctuations at $ T=0 $ and
   makes the RAFHM dramatically different than the RFHM \cite{rh}.

\section{ $ 1/S $ corrections to the Linear spin wave results. }

Normal-ordering $\mathcal{H}_4$ in the quasi-particle $\alpha, \beta$ basis
in Eq.\ref{LSW} ( namely with respect to the quantum ground state $ | \Omega \rangle $ in the next section )
results in the three terms:
\begin{align}
    \mathcal{H}_4
    =\mathcal{H}_4^{(0)}
     +\mathcal{H}_4^{(2)}
      +\mathcal{H}_4^{(4)}
\end{align}
where $\mathcal{H}_4^{(0)}$ is a constant term,
$\mathcal{H}_4^{(2)}$ is quadratic and $\mathcal{H}_4^{(4)}$ is quartic in terms of $ \alpha, \beta $.
Following \cite{japan}, it is easy to see that  $\mathcal{H}_4^{(0)}$ and $\mathcal{H}_4^{(2)}$ contribute to the ground state energy and
energy spectrum at the order of $ 1/S $ corrections to the LSWE respectively. While $\mathcal{H}_4^{(4)}$ only make contributions to
higher order than $ 1/S^2 $, so can be dropped at the order of $ 1/S $.

   The $ \mathcal{H}_4^{(0)} $ leads to the $1/S $ corrections ( in unit $ 2JS $ ) to the ground state energy listed in Eqn.\ref{eM}:
\begin{align}
    \mathcal{H}_4^{(0)}=-\frac{N}{8S}[I_0^2+I_c^2+I_s^2]
\label{H40}
\end{align}
    where
\begin{eqnarray}
    I_0(\beta)
   & = &\int\frac{d^2 k}{4\pi^2}
    \Big[
	\cos k_x\sin\theta_k\Big(
	\frac{\gamma_k^-}{\omega_k^-}-\frac{\gamma_k^+}{\omega_k^+}
	\Big)
	+\frac{1}{\omega_k^-}+\frac{1}{\omega_k^+}-2
    \Big],\label{I0}     \nonumber  \\
    I_c(\beta)
   & = &\int\frac{d^2k}{4\pi^2}
    \Big[
	-\cos k_y\Big(
	\frac{\gamma_k^-}{\omega_k^-}+\frac{\gamma_k^+}{\omega_k^+}
	\Big)
	+\Big(\frac{1}{\omega_k^-}+\frac{1}{\omega_k^+}-2\Big)
	\cos2\beta
    \Big],\label{I1}    \nonumber  \\
    I_s(\beta)
   & = &\int\frac{d^2 k}{4\pi^2}
    \Big[
	\sin k_y\cos\theta_k\Big(
	\frac{\gamma_k^-}{\omega_k^-}-\frac{\gamma_k^+}{\omega_k^+}
	\Big)
	+\Big(\frac{1}{\omega_k^-}+\frac{1}{\omega_k^+}-2\Big)
	\sin2\beta
    \Big],\label{I2}
\end{eqnarray}
    The numerical results of Eqn.\ref{H40} for $ S=1/2 $  was drawn in Fig.\ref{eM1s}a.
    There is always  $1/S$ correction to the ground state energy shown in Fig.\ref{eM1s}a, but found to be small.

    The $  \mathcal{H}_4^{(2)} $ can be written as in the normal ordered form:
\begin{align}
	\mathcal{H}_4^{(2)}=-\frac{1}{4S}\sum
	\big[
		&2(I_0+I_c\cos2\beta+I_s\sin2\beta)
		(A_k^\dagger A_k+A_k^\dagger C_{-k}^\dagger
		+B_k^\dagger B_k+B_k^\dagger D_{-k}^\dagger+h.c.)\nonumber\\
		&+2I_0\cos k_x
		(A_{-k}A_k+C_{-k}^\dagger A_{-k}
	              +C_{k}^\dagger A_k+C_{k}^\dagger C_{-k}^\dagger
		+B_{-k}B_k+D_{-k}^\dagger B_{-k}
		+D_{k}^\dagger B_k+D_{k}^\dagger D_{-k}^\dagger+h.c.)\nonumber\\
		&+2(I_c\cos k_y-I_s\sin k_y)
		(A_{-k}B_k+D_{-k}^\dagger A_{-k}+C_{k}^\dagger B_k+C_{k}^\dagger D_{-k}^\dagger+h.c.)
	\big]
\label{H42}
\end{align}
where $ A, B, C, D $ are the annihilation operators in terms of $ \alpha, \beta $:
\begin{align*}
	A_k&=\sin(\theta_k/2)\cosh\phi_k^-\alpha_k
	     +\cos(\theta_k/2)\cosh\phi_k^+\beta_k,\qquad
	B_k=-\cos(\theta_k/2)\cosh\phi_k^-\alpha_k
	      +\sin(\theta_k/2)\cosh\phi_k^+\beta_k\\
	C_{-k}&=\sin(\theta_k/2)\sinh\phi_k^-\alpha_{-k}
	    	+\cos(\theta_k/2)\sinh\phi_k^+\beta_{-k},\quad
	D_{-k}=-\cos(\theta_k/2)\sinh\phi_k^-\alpha_{-k}
	      	 +\sin(\theta_k/2)\sinh\phi_k^+\beta_{-k}.
\end{align*}
  Following \cite{japan},  it is straightforward to evaluate the $ 1/S $ correction the spectrum Eq.\ref{LSW} obtained at the LSW,
  which, in turn, changes the gap and leads to the shifts of $ \beta_1, \beta_2 $, the minimum positions $ (0, k^{0}_y ) $ of the $ C-IC $ relativistic magnons  as shown in Fig.\ref{groundex}. We also evaluate its contribution to the magnetization shown in Fig.\ref{eM1s}b.
  They are all
  At $ \beta=\pi/4 $, in a suitably chosen rotated basis, we find $ \mathcal{H}_4^{(2)} $ can be written as $
  \mathcal{H}_4^{(2)}= C \sum_\mathbf{k}
	(\omega_\mathbf{k}^-\alpha_\mathbf{k}^\dagger\alpha_\mathbf{k}
	+\omega_\mathbf{k}^+\beta_\mathbf{k}^\dagger\beta_\mathbf{k}) $ which just contributes to a multiple factor to the LSW spectrum
in Eq.\ref{LSW}, so it does not change the magnetization in Fig.\ref{eM1s}b at the order of $1/S $.
There is no $ 1/S $ corrections to the magnetization at the two Abelian points $ \beta=0,\pi/2 $,
consistent with the results achieved at the AFM Heisenberg model \cite{japan}.
Of course, it does not shift the minima positions $ (0, k^{0}_y = \pm \pi/2 ) $ in Fig.\ref{groundex}a as dictated by the Mirror symmetry.
It does lead to a multiple factor to the gap at the order of $1/S $ shown in Fig.\ref{groundex}b.
There are $1/S $ corrections at any other $ \beta $ shown in Fig.\ref{eM1s}b, but found to be small even at $ S=1/2 $.
We expect that the magnetization will receive $ 1/S^2 $ corrections at $ \beta=0, \pi $ as calculated in \cite{japan} and also at $ \beta=\pi/4 $.


Note that the $ (\beta_1, \beta_2) $ in Fig.\ref{groundex}a take the same values as those in
the RFHM at the order $ 1/S $. This is because, as shown in the last section, the two models share the same
unitary transformation Eqn.\ref{unit}. However, the main difference is that the RAFHM also involves a
Bogoliubov transformation Eqn.\ref{bog} which induces quantum
fluctuations at $ T=0 $. In sharp contrast, the $Y-x$ ground state is exact for the RFHM \cite{rh},
there are no quantum corrections to any orders in $ 1/S $ at $ T=0 $.
So in the RFHM, $ \beta_1 $ is exact, will not receive any quantum corrections from higher order
expansions in $ 1/S $. While, in the RAFHM, $ \beta_1 $ is not exact, does receive  quantum corrections from higher order
expansions in $ 1/S $. The contributions at the $1/S $ order is shown in Fig.\ref{groundex}a and found to be very small.

Note that the Mirror symmetry dictates that (1) the minima positions in Fig.\ref{groundex}a is exactly symmetric about $ \beta=\pi/4 $.
(2) The minimum position at $ \beta=\pi/4 $ is exactly pinned at $ (0, k^{0}_y=\pi/2 ) $
(3) The relation $ \beta_2= \pi/2- \beta_1 $ is exact.
All these 3 exact statements should receive no corrections
to any orders in $1/S $. Indeed, we find they receive no correction at the $1/S $ order.

\section{ Quantum corrections to the classical state and the $ U(1)_{soc} $ symmetry of the quantum ground state to the order $ 1/S $. }

{\sl 1. Quantum corrections to the classical Y-y state }

The $ Y-x $ state is the exact ground state of the RHFM \cite{rh}, so no quantum fluctuations.
However, the $ Y-y $ state in Fig.\ref{u1}a is only classical, valid only in the $ S= \infty $.
Any finite $ S $ causes quantum corrections to the classical ground state.
The quantum corrections to the Halperin's $ (1,1,1) $ state in the bilayer or trilayer quantum Hall state due to the neutral gapless Goldstone mode
was investigated in \cite{blqh,tri}. In fact, the $ (1,1,1) $ state is a ferromagnetic state which is exact only when distance between the two layers
vanishes. At any finite distance, it suffers quantum fluctuations and should receive quantum corrections.
Similar quantum corrections to the classical Bose-Einstein condensation (BEC) $ \langle \Psi \rangle= a \neq 0 $
in the superfluid Helium can also be evaluated.

   Using the fact that $ | \Omega \rangle $  is the vacuum of the quasi-particle operators: $ \alpha_k | \Omega \rangle = \beta_k | \Omega \rangle =0 $,
   we find the quantum fluctuations corrected ground state at the order $1/S $:
\begin{align}
	|\Omega\rangle=\mathcal{C}\exp\left\{
	\sum_k \Big[\sin\theta_k
	\Big(
	\frac{1-\omega_k^-}{2\gamma_k^-}+\frac{\omega_k^+-1}{2\gamma_k^+}
	\Big)
	(a_k^\dagger a_{-k}^\dagger+b_k^\dagger b_{-k}^\dagger)
	+2\sin^2\Big(\frac{\theta_k}{2}\Big)
	\Big(
	\frac{\omega_k^--1}{\gamma_k^-}+\frac{\omega_k^+-1}{\gamma_k^+}
	\Big)
	a_k^\dagger b_{-k}^\dagger
	\Big]
	\right\}
	|Y-y\rangle
\label{ground}
\end{align}
   which establishes the connection between the quantum ground state $  |\Omega\rangle $ and
   the classical ground state $ |Y-y\rangle $.
   Obviously, $ |Y-y\rangle $  is the vacuum of the original boson operators $ a $ and $ b $, while
   $ |\Omega\rangle $ is that of the quasi-particle operators  $ \alpha_k $ and $ \beta_k $ which
   contain all the information of the quantum fluctuation generated $ C-C_0 $ and $ C-IC $ magnons.

{\sl 2. The $ U(1)_{soc} $ symmetry of the quantum ground state }

In the classical limit, we know $ Q_c| Y-y \rangle=0$ where $ Q_c= \sum_i (-1)^{i_x} S^{y}_i $ is the conserved quantity along the line.
Here we show that in the strong coupling limit, the quantum fluctuations corrected ground state $ | \Omega \rangle $ also
satisfies  $ Q_c | \Omega \rangle =0 $ at the order $1/S $.

For the notational convenience, we apply a globe rotation
$(S_i^x,S_i^y,S_i^z)\to (\tilde{S}_i^x,\tilde{S}_i^z,-\tilde{S}_i^y)$
to rotate $ S^y $ to $ \tilde{S}_i^z $, then the
conserved quantity takes the form
\begin{equation}
    Q_c  =  \sum_i(-1)^{i_x}S_i^y  =\sum_i(-1)^{i_x}\tilde{S}_i^z
     =  -\sum_k(a_k^\dagger a_{k+Q_x}-b_k^\dagger b_{k+Q_x})
\end{equation}
where $Q_x=(\pi,0)$ is the orbital structure of conserved quantity $Q_c$.

  Combining Eqn.\ref{unit} and Eqn.\ref{bog} lead to:
\begin{align}
	a_k^\dagger
	&=\sin(\theta_k/2)\bar{a}_k^\dagger
	+\cos(\theta_k/2)\bar{b}_k^\dagger
	=\sin(\theta_k/2)(u_k^a\alpha_k^\dagger+v_k^a\alpha_{-k})
	+\cos(\theta_k/2)(u_k^b\beta_k^\dagger+v_k^b\beta_{-k})   \nonumber  \\
	b_k^\dagger
	&=\sin(\theta_k/2)\bar{b}_k^\dagger
	-\cos(\theta_k/2)\bar{a}_k^\dagger
	=\sin(\theta_k/2)(u_k^b\beta_k^\dagger+v_k^b\beta_{-k})
	-\cos(\theta_k/2)(u_k^a\alpha_k^\dagger+v_k^a\alpha_{-k})
\end{align}
     Using $ \alpha_k | \Omega \rangle = \beta_k | \Omega \rangle =0 $ and the bosonic commutation relations of $ \alpha_k, \beta_k $
     simplifies it to:
\begin{align}
	a_k^\dagger a_{k+Q_x}| \Omega \rangle
	&=[\sin(\theta_k/2)u_k^a\alpha_k^\dagger
	+\cos(\theta_k/2)u_k^b\beta_k^\dagger]
	[\sin(\theta_{k+Q_x}/2)v_{k+Q_x}^a\alpha_{-k-Q_x}^\dagger
	+\cos(\theta_{k+Q_x}/2)v_{k+Q_x}^b\beta_{-k-Q_x}^\dagger]|\Omega\rangle    \nonumber  \\
	b_k^\dagger b_{k+Q_x}| \Omega \rangle
	&=[\sin(\theta_k/2)u_k^b\beta_k^\dagger
	-\cos(\theta_k/2)u_k^a\alpha_k^\dagger]
	[\sin(\theta_{k+Q_x}/2)v_{k+Q_x}^b\beta_{-k-Q_x}^\dagger
	-\cos(\theta_{k+Q_x}/2)v_{k+Q_x}^a\alpha_{-k-Q_x}^\dagger]| \Omega\rangle
\end{align}
thus
\begin{eqnarray}
	Q_c|\Omega \rangle
	& = & \sum_k\big(
	\cos[(\theta_k+\theta_{k+Q_x})/2]u_k^av_{k+Q_x}^a\alpha_k^\dagger\alpha_{-k-Q_x}^\dagger
	-\cos[(\theta_k+\theta_{k+Q_x})/2]u_k^bv_{k+Q_x}^b\beta_k^\dagger\beta_{-k-Q_x}^\dagger   \nonumber  \\
	&-& \sin[(\theta_k+\theta_{k+Q_x})/2]u_k^av_{k+Q_x}^b\alpha_k^\dagger\beta_{-k-Q_x}^\dagger
	-\sin[(\theta_k+\theta_{k+Q_x})/2]v_{k+Q_x}^au_k^b\alpha_{-k-Q_x}^\dagger\beta_k^\dagger\big)| \Omega \rangle
\end{eqnarray}
  From Eqn.\ref{unit}, one can see $\theta_k+\theta_{k+Q_x}=0$, so the above equation is simplified to:
\begin{eqnarray}
	Q_c| \Omega  \rangle
	& =  &\sum_k u_k^av_{k+Q_x}^a\alpha_k^\dagger\alpha_{-k-Q_x}^\dagger| \Omega\rangle
	-\sum_k u_k^bv_{k+Q_x}^b\beta_k^\dagger\beta_{-k-Q_x}^\dagger| \Omega\rangle     \nonumber  \\
	&=&\sum_k \frac{1}{2}(u_k^av_{k+Q_x}^a+u_{k+Q_x}^av_k^a)
	\alpha_k^\dagger\alpha_{-k-Q_x}^\dagger| \Omega \rangle
	-\sum_k \frac{1}{2}(u_k^bv_{k+Q_x}^b+u_{k+Q_x}^bv_k^b)
	\beta_k^\dagger\beta_{-k-Q_x}^\dagger| \Omega  \rangle
\end{eqnarray}
  From Eqn.\ref{bog}, one can see $\lambda_{k+Q_x}^\pm=-\lambda_{k}^\pm$ and $\omega_{k+Q_x}^\pm=\omega_{k}^\pm$,
  which lead to $Q_c| \Omega  \rangle=0$  at the order of $ 1/S $.
  Of course, it should hold exactly, so to any order of $ 1/S $.

  Although the classical $ Y-y $ state contains no information on the $ C-C_0, C-IC $ relativistic magnons, the quantum ground state $ | \Omega \rangle $
  does contain it and can be detected by the transverse structure factor Eq.\ref{Sfactor} precisely.

\section{ The fermionic excitations in  the Y-y state at weak coupling $ U \ll t $ }

    The two branches of gapped fermionic excitations in the Y-y state at weak coupling are:
\begin{align}
	\epsilon_1
	&=2t \sqrt{\sin^2 k_x\!+\!\cos^2\beta\cos^2 k_y\!+\!\sin^2\beta\sin^2 k_y
	\!+\frac{M^2}{16 t^2}
	\!-\!2\sqrt{\cos^2\beta\cos^2 k_y\!
		   (\sin^2 k_x+\!\sin^2\beta\sin^2 k_y)}}   \nonumber  \\
	\epsilon_2
	&=2t \sqrt{\sin^2 k_x\!+\!\cos^2\beta\cos^2 k_y\!+\!\sin^2\beta\sin^2 k_y
	\!+\frac{M^2}{16 t^2}
	\!+\!2\sqrt{\cos^2\beta\cos^2 k_y\! (\sin^2 k_x + \!\sin^2\beta\sin^2 k_y)}}
\label{fer}
\end{align}
where $ M \sim e^{-3/[4U\rho_0(\beta)]} $ where $ \rho_0(\beta) $ is the DOS
at the FS in Fig.\ref{squaredash}a with the asymptotic behavior
$ \rho_0(\beta) \sim \ln 1/\beta $ when $ \beta \rightarrow 0 $ and $ \rho_0(\beta) \rightarrow 0 $
when $ \beta \rightarrow \pi/2^{-} $.







\end{widetext}



\begin{thebibliography}{99}


\bibitem{wen0} J. R. Schrieffer, X. G. Wen, and S. C. Zhang,
Dynamic spin fluctuations and the bag mechanism of high-T, superconductivity,
Phys. Rev. B \textbf{39}, 11663 (1989).


\bibitem{scaling} A. V. Chubukov,  S. Sachdev, and  J. Ye,
Theory of two-dimensional quantum Heisenberg antiferromagnets with a nearly critical ground state,
Phys. Rev. B \textbf{49}, 11919(1994).




\bibitem{aue} A. Auerbach,
Interacting electrons and quantum magnetism,
(Springer Science \& Business Media, 1994).

\bibitem{wenbook} X.G. Wen,  Quantum Field Theory of Many-body Systems, From the Origin of Sound
to an Origin of Light and Electrons, ( OXFORD UNIVERSITY PRESS, 2004  ),

\bibitem{sachdev} S. Sachdev, Quantum Phase transitions, (2nd edition, Cambridge University Press, 2011).


\bibitem{stevermp}  S. A. Kivelson, {\sl et.al},
How to detect fluctuating stripes in the high-temperature superconductors, Rev. Mod. Phys. 75, 1201 (2003).



\bibitem{rashba} Y. A. Bychkov and E.I. Rashba, J. Phys. C 17, 6039 (1984)

\bibitem{ahe} Jinwu Ye, Yong Baek Kim, A. J. Millis, B. I. Shraiman, P. Majumdar, and Z.
 Tesanovic  Berry phase theory of the Anomalous Hall Effect: Application to Colossal Magnetoresistance Manganites, Phys. Rev. Lett. 83, 3737 (1999)

\bibitem{socsemi} Lev P. Gor'kov and Emmanuel I. Rashba, Superconducting 2D System with Lifted Spin Degeneracy: Mixed Singlet-Triplet State,
Phys. Rev. Lett. 87, 037004 (2001).

\bibitem{ahe2}  T. Jungwirth, Qian Niu and A. H. MacDonald, Anomalous Hall Effect in Ferromagnetic Semiconductors,   Phys. Rev. Lett. 88, 207208 (2004).

\bibitem{she}  J. Sinova, {\sl et.al}, Universal intrinsic spin Hall effect, Phys. Rev. Lett. 92, 126603 (2004).

\bibitem{niu}  Wang Yao and Qian Niu, Berry Phase Effect on the Exciton Transport and on the Exciton Bose-Einstein Condensate, Phys. Rev. Lett. 101, 106401 (2008).

\bibitem{aherev}  Naoto Nagaosa, Jairo Sinova, Shigeki Onoda, A. H. MacDonald, and N. P. Ong, Anomalous Hall effect,
Rev. Mod. Phys. 82, 1539 (2010) - Published 13 May 2010.


\bibitem{sherev}  Jairo Sinova, Sergio O. Valenzuela, J. Wunderlich, C.?H. Back, and T. Jungwirth, Spin Hall effects,
Rev. Mod. Phys. 87, 1213 (2015) - Published 27 October 2015













\bibitem{blochrmp}  Immanuel Bloch, Jean Dalibard, and Wilhelm Zwerger,  Many-body physics with ultracold gases,
Rev. Mod. Phys. 80, 885 (2008)

\bibitem{coldafm} Russell A. Hart, Randall G. Hulet {\sl et.al},
Observation of antiferromagnetic correlations in the Hubbard model with ultracold atoms, Nature 519, 211每214 (12 March 2015).



\bibitem{expk40} Lianghui Huang, {\sl et.al},
Experimental realization of a two-dimensional synthetic spin-orbit coupling in ultracold Fermi gases, Nature Physics 12, 540-544 (2016).

\bibitem{expk40zeeman} Zengming Meng, {\sl et.al},
Experimental observation of topological band gap opening in ultracold Fermi gases with two-dimensional spin-orbit coupling, arXiv:1511.08492.

\bibitem{clock} Michael L. Wall, {\sl et.al}, Synthetic Spin-Orbit Coupling in an Optical Lattice Clock, Phys. Rev. Lett. 116, 035301 (2016).

\bibitem{2dsocbec} Zhan Wu, {\sl et.al}, Realization of Two-Dimensional Spin-orbit Coupling for Bose-Einstein Condensates, Science 354, 83-88 (2016).

\bibitem{notation} Here we still use the same notation used in \cite{rh}. In the Y-$(0,\pi) $ called Y-y state, the first letter indicates
the spin polarization, the second letter indicates the orbital order. In the $C-C_0 $ magnons, the first letter indicates
the ground state is  Commensurate, the second letter indicates the excitation is also commensurte  with  its minimum at $\boldmath{k}=0 $.
In the C-IC magnons, the first letter indicates
the ground state is still Commensurate, the second letter indicates the excitation is In-commensurte
with its  minimum at in-commensurate momenta $ \boldmath{k}=( 0, \pm k^0_y) $.
By the spin-orbital coupling (SOC) in Eq.\ref{rh} which, in fact, is a spin only model
we mean the spin-bond coupling, namley, the spin-spin exchange interaction is bond-dependent in the form of
the two $ SO(3) $ rotation matrix $ R(X, 2 \alpha),  R(Y, 2 \alpha) $ in Eq.\ref{rh}. For example, the celebrated Kitaev model
has 3 bond-dependant spin-spin interactions.

\bibitem{kondo1} Jinwu Ye, On Emery-Kivelson line and universality of Wilson ratio of
  spin anisotropic Kondo model, Phys. Rev. Lett. 77, 3224 (1996).

\bibitem{kondo2} Jinwu Ye, Abelian Bosonization approach to quantum impurity problems, Phys. Rev. Lett. 79, 1385 (1997)

\bibitem{dimer}  Daniel S. Rokhsar and Steven A. Kivelson, Superconductivity and the Quantum Hard-Core Dimer Gas, Phys. Rev. Lett. 61, 2376 每 Published 13 November 1988.

\bibitem{dimer2} Hong Yao and Steven A. Kivelson, Exact Spin Liquid Ground States of the Quantum Dimer Model on the Square and Honeycomb Lattices,
Phys. Rev. Lett. 108, 247206 (2012) 每 Published 13 June 2012.

\bibitem{rh} Fadi Sun, Jinwu Ye, and Wu-Ming Liu,
Quantum magnetism of spinor bosons in optical lattices with synthetic non-Abelian gauge fields,
Phys. Rev. A \textbf{92}, 043609 (2015).

\bibitem{rhrashba}   Fadi Sun, Jinwu Ye, Wu-Ming Liu, Strongly interacting spinor bosons with Rashba spin-orbital couplings on  a square lattice,  arXiv:1603.00451



\bibitem{sw1}
Scalettar, R. T., Batrouni, G. G., Kampf, A. P. \& Zimanyi, G. T.
Simultaneous diagonal and off-diagonal order in the Bose-Hubbard Hamiltonian.
\textit{Phys. Rev. B} \textbf{51}, 8467 (1995).

\bibitem{sw2}
Murthy, G., Arovas, D. \& Auerbach, A.
Superfluids and supersolids on frustrated two-dimensional lattices.
\textit{Phys. Rev. B} \textbf{55}, 3104 (1997).

\bibitem{japan}  Jun-ichi Igarashi, 1/S expansion for thermodynamic quantities in a two-dimensional Heisenberg antiferromagnet at zero temperature,
Phys. Rev. B 46, 10763每10771 (1992); Jun-ichi Igarashi and Tatsuya Nagao, 1決S-expansion study of spin waves in a two-dimensional Heisenberg antiferromagnet,
Phys. Rev. B 72, 014403 (2005).


\bibitem{higgs} Yu Yi-Xiang, Ye Jinwu and  Liu W. M,
Goldstone and Higgs modes of photons inside an cavity and their detections.
\textit{Scientific Reports} \textbf{3}, 3476 (2013).

\bibitem{setting}
At $ \beta_1,\beta_2 $, from $ v_y \sim |\beta-\beta_i|^{1/2} $ and a simple scaling analysis,
one can just set $ v_y \sim T^{1/4} $ in all the physical quantities in Eqn.\ref{CmT} and \ref{pizero}.
The Wilson ratio stays the same.


\bibitem{tqpt}
Sun, F., Yu, X.-L., Ye, J., Fan, H. \& Liu, W.-M.
Topological Quantum Phase Transition in Synthetic Non-Abelian Gauge Potential: Gauge Invariance and Experimental Detections.
\textit{Sci. Rep.} \textbf{3}, 2119 (2013).
As stressed in this work, in contrast to condensed matter systems where only gauge invariant quantities can be measured,
both gauge invariant and non-invariant quantities can be measured in cold atom systems by experimently generating various non-abelian gauges
corresponding to the same set of Wilson loops. See also \cite{rh}.





\bibitem{jin} Stewart, J. T., Gaebler, J. P. and Jin, D. S.
Using photoemission spectroscopy to probe a strongly interacting Fermi gas. Nature, 454, 744 (2008).

\bibitem{lightatom1}
Ye, J. \textit{et al}.
Light-scattering detection of quantum phases of ultracold atoms in optical lattices.
\textit{Phys. Rev. A} \textbf{83}, 051604 (2011).

\bibitem{lightatom2}
Ye, J., Zhang, K. Y., Li, Y., Chen, Y. \& Zhang, W. P.
Optical Bragg, atom Bragg and cavity QED detections of quantum phases and excitation spectra of ultracold atoms in bipartite and frustrated optical lattices.
\textit{Ann. Phys.} \textbf{328}, 103 (2013).


\bibitem{heat1}
Kinast, J. \textit{et al}.
Heat Capacity of a Strongly Interacting Fermi Gas.
\textit{Science} \textbf{307}, 1296 (2005).

\bibitem{heat2}
Ku, M. J. H. \textit{et al}.
Revealing the Superfluid Lambda Transition in the Universal Thermodynamics of a Unitary Fermi Gas.
\textit{Science} \textbf{335}, 563 (2012).

\bibitem{dosexp} Gemelke, N., Zhang X., Huang C. L., and Chin, C. In situ observation of incompressible Mott-insulating domains in ultracold atomic gases, Nature (London) \textbf{460}, 995 (2009).

\bibitem{blqh} Longhua Jiang and Jinwu Ye,
Ground state, quasihole and a pair of quasihole wavefunctions in Bi-layer Quantum Hall systems, Phys. Rev. B 74, 245311 (2006).

\bibitem{tri} Jinwu Ye,  Broken symmetry, excitons, gapless modes, and topological excitations in trilayer quantum Hall systems,
Phys. Rev. B 71, 125314 (2005).

\bibitem{bosonmsg} L. Balents, L. Bartosch, A. Burkov, S. Sachdev, K. Sengupta, Phys. Rev. B 71 (2005) 144508.

\bibitem{honey1}
Jiang, L. \& Ye, J.
The mobility of dual vortices in honeycomb, square, triangular, Kagome and dice lattices.
\textit{J. Phys, Condens. Matter} \textbf{18}, 6907 (2006).

\bibitem{dual1}
Ye, J.
Duality, magnetic space group and their applications to quantum phases and phase transitions on bipartite lattices in several experimental systems.
\textit{Nucl. Phys. B} \textbf{805}, 418 (2008).

\bibitem{dual2} Chen, Y. \& Ye, J., Characterizing boson orders in lattices by vortex degree of freedoms.
\textit{ Philos. Mag. } \textbf{92}, 4484-4491 (2012).

\bibitem{dual3}
Ye, J. \& Chen, Y.
Quantum phases, Supersolids and quantum phase transitions of interacting bosons in frustrated lattices.
\textit{Nucl. Phys. B} \textbf{869}, 242 (2013).

\bibitem{cwu}
Zi Cai, Xiangfa Zhou, and Congjun Wu, Magnetic phases of bosons with synthetic spin-orbit coupling in optical lattices,
Phys. Rev. A 85, 061605(R), 2012.

\bibitem{classdm1} J. Radic∩, A. Di Ciolo,  K. Sun, and V. Galitski, Exotic Quantum Spin Models in Spin-Orbit-Coupled Mott Insulators, PRL 109, 085303 (2012)

\bibitem{classdm2} William S. Cole1, Shizhong Zhang, Arun Paramekanti, and Nandini Trivedi,
Bose-Hubbard Models with Synthetic Spin-Orbit Coupling: Mott Insulators, Spin Textures, and Superfluidity, Phys. Rev. Lett. 109, 085302 (2012) [5 pages].



\bibitem{rhh} Fadi Sun, Jinwu Ye, Wu-Ming Liu, Quantum incommensurate Skyrmion  crystal phases and Commensurate to In-commensurate transitions of cold atoms and materials with spin orbit couplings, arXiv:1502.05338.

\bibitem{rhht} Fadi Sun,  Jinwu Ye and Wu-Ming Liu,
Classification of magnons in Rotated Ferromagnetic Heisenberg model and their competing responses in transverse fields, Phys. Rev. B 94, 024409 每 Published 7 July 2016.





\bibitem{ssdw} Shang-Shun Zhang, Jinwu Ye, Wu-Ming Liu,
 Itinerant magnetic phases and quantum Lifshitz transitions in repulsively interacting spin-orbit coupled Fermi gas, Phys. Rev. B 94, 115121 每 Published 9 September 2016.

\bibitem{pairing} Yi-Xiang Yu, Jinwu Ye, Wu-Ming Liu, Cherence length in attractively interacting Fermi gases with Spin-orbit Couplings,
Phys. Rev. A 90, 053603  (2014).


\bibitem{so5} Congjun Wu, Jiang-ping Hu, and Shou-cheng Zhang, Exact SO(5) Symmetry in the Spin-3/2 Fermionic System,
Phys. Rev. Lett. 91, 186402 (2003).











\end{thebibliography}
\end{document}